\documentclass{article}
\usepackage{lipsum}
\usepackage{authblk}
\usepackage[english]{babel}
\usepackage{amsfonts}
\usepackage{amsmath}
\usepackage[top=2cm, bottom=2cm, left=2cm, right=2cm]{geometry}
\usepackage{fancyhdr}
\usepackage{multicol}
\usepackage{epsfig}
\usepackage{graphics}
\usepackage{epsfig}
\usepackage{lipsum}
\usepackage{fancyhdr}

\renewenvironment{abstract}{
	
	\hfill\begin{minipage}{0.95\textwidth}
		\rule{\textwidth}{1pt}}
		{\par\noindent\rule{\textwidth}{1pt}\end{minipage}
	}

\newcommand{\bra}[1]{\bigl\langle #1 \bigr|}
\newcommand{\ket}[1]{\bigl| #1 \bigr\rangle}


\begin{document}
%
\title{\textbf{Bidirectional Teleportation using Fisher Information}}
\author[1]{ \textbf{C. Seida}}
\author[2,3]{ \textbf{A. El Allati}}
\author[4,5]{ \textbf{N. Metwally}}
\author[1]{ \textbf{Y. Hassouni}}
\affil[1]{\small ESMAR, Mohammed V University, Faculty of Sciences, Rabat, Morocco}
\affil[2]{\small Laboratory of R\&D in Engineering Sciences, Faculty of Sciences and Techniques Al-Hoceima, Abdelmalek Essaadi University, T\'{e}touan,
	Morocco}
\affil[3]{\small  The Abdus Salam International Center for Theoretical Physics,
	Strada Costiera 11, Miramare-Trieste, Italy}
\affil[4]{\small Math. Dept.,
	College of Science, University of Bahrain,  P. O. Box 32038
	Kingdom of Bahrain}
\affil[5]{\small Math. Dept. Faculty of
	Science, Aswan University, Aswan, Egypt}
	
\maketitle
\begin{abstract}
In this contribution, we reformulated the bidirectional teleportation protocol suggested in \cite{Kiktenko}, by means of Bloch vectors as well as the local operations are represented by using Pauli operators. Analytical and numerical calculations for the teleported state and Fisher information are introduced.  It is shown that both quantities depend on the initial state settings of the teleported qubits and their triggers. The Fidelities and the Fisher information of the bidirectionally teleported states are maximized when the qubit and its trigger are polarized in the same direction. The minimum values are predicted if both initial qubits  have different polarization or non-zero phase. The maximum values of the Fidelity and the quantum Fisher information are the same, but they are predicted at different polarization angles. We display that the multi-parameter form is much better than the single parameter form, where  it satisfies the bounds of classical, entangled systems and the uncertainty principle.

\end{abstract}

\section{Introduction}\label{section1}
Quantum teleportation is a process to transmit quantum information between two distant partners,  Alice and Bob by using a pre-shared entanglement and classical communication \cite{1}. Over the last twenty
years, many modified protocols of quantum teleportation have been proposed, such as the bidirectional quantum
teleportation $(BQT)$ and the controlled bidirectional quantum teleportation $(CBQT)$ protocols. Moreover, these protocols  are designed to
transfer quantum states simultaneously between the two users. The  $CBQT$ protocols necessitate an intervention from  a third part (Charlie) to accomplish the task.  In 2001, $Huelga$ $et$ $al.$ \cite{2} discussed the remote implementation of non-local unitary gates using the teleportation in opposite sides, which is bidirectional quantum teleportation. $Zha$ $et$ $al.$ \cite{Zha} proposed a controlled bidirectional teleportation scheme using five qubit cluster state.  $Lie$ $et$ $al.$ \cite{Lie} suggested a scheme by using GHZ-Bell composite state. $Zadeh$ $et$ $al.$ \cite{Zadeh} presented a BQT of two-qubit states by using eight qubits entangled state. $Choudhury$ $et$ $al.$ \cite{Choudhury} proposed a protocol
based on a quantum channel of eleven qubits entangled state to transmit three-qubit $W$ state. In the context of non-perfect Teleportation, $Kiktenko$ $et$ $al.$ \cite{Kiktenko} examined a non-perfect protocol of BQT where Alice and Bob transfer a noisy version of their states to each other using a single bell state as a quantum channel.\\

In some  situations, it is not needed to teleport  the global information of the state, but only the information of one parameter or multiparameter of the system. The QFI  plays a primordial role in the estimation of an unknown parameter through the Cramer-Rao bound \cite{paris,Helstrom,Holevo}. Furthermore, the quantum Fisher information matrix QFIM, provides a tool to determine the precision in any multiparametric estimation protocol. Besides the quantum estimation, the QFI and the QFIM are crucial concepts in Quantum metrology \cite{Giovannetti,Giovannetti1}, quantum phase transition\cite{Wu,Marzolino} and entanglement detection \cite{Luo}. Many efforts have been devoted to investigate the behavior of the QFI in different systems. Metwally $et$ $al.$ \cite{Metwally} estimated the teleported and the gained parameters by means of Fisher information in a non-inertial frame. El Anouz $et$ $al.$ \cite{El Anouz} discussed the unidirectional teleportation of QFI in one and two-qubit states. Xiao $et$ $al.$ \cite{Xiao} studied the QFI teleportation under decoherence by utilizing the partial measurements. Jafarzadeh $et$ $al.$ \cite{Jafarzadeh} investigated how thermal noise affects the quantum resources and quantum Fisher information (QFI) teleportation.\\

In the present work, we consider two legitimates users who can transfer their states to each other following $Kiktenko$ protocol \cite{Kiktenko}. We use a single Bell state as a quantum channel to teleport the parameters of the initial states in two opposite directions. The Fidelity of the Bidirectional teleported states between users is discussed, and the estimation of weight parameters of the teleported states Bidirectionally is quantified using QFI. We show that, the QFI and the Fidelity of the bidirectional teleportation depend on the initial states of the teleported qubits as well as on the states of the Triggers. Where, the maximum values of the Fidelity and the quantum Fisher information are the same, but they are predicted at different polarization angles. Also, we display that the multi-parameter estimation  is much better than the single parameter form.\\

The paper is structured as follows. In Sec.\ref{section2} we briefly review the bidirectional quantum teleportation protocol of  $Kiktenko$ $et$.$al$.\cite{Kiktenko} where we described it by using Bloch vector and Pauli operators representation. The Fidelities of the bidirectional teleportation states are investigated for different initial state settings in Sec.\ref{section2.5}. The mathematical formals of the quantum Fisher information for a single and multi-parameter are described in Sec.3. Finally, we summarize our results in Sec.\ref{section4}.

\section{Bidirectional teleportation protocol}\label{section2}
In this framework, we use a protocol proposed by Kiktenko $et$ $al.$ \cite{Kiktenko} to transfer information from Bob to Alice as well as from Alice to Bob, i.e.,  bidirectional teleportation. This protocol is based on a single Bell state, two trigger qubits and  classical communication in both directions.

\subsection{Preparing the initial states}
\begin{figure}[h!]
	\begin{center}
		\includegraphics[scale=.45]{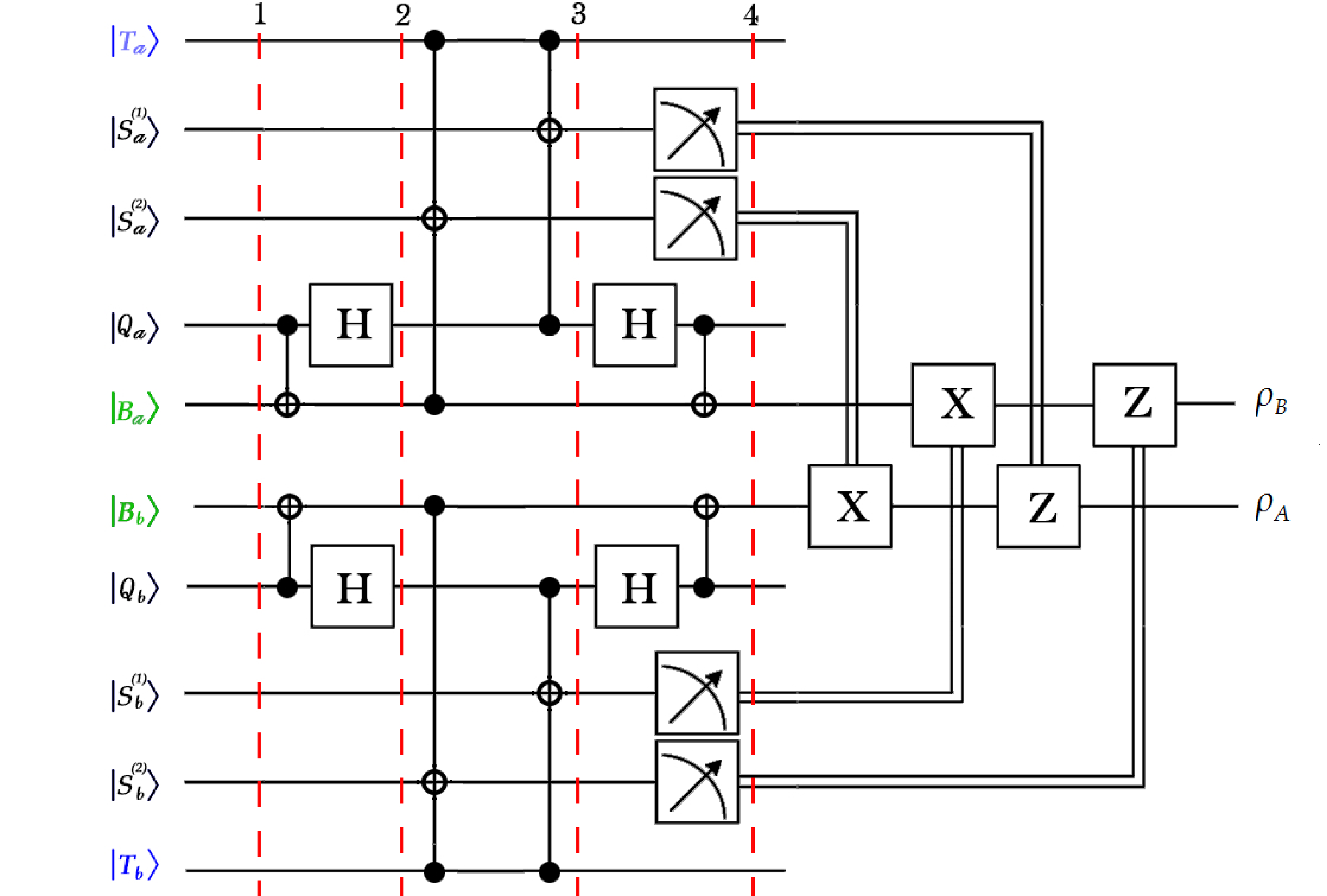}
		\caption{Quantum circuit of the bidirectional teleportation between Alice and Bob using a single Bell state (Eq. (\ref{bell})), two trigger qubits $\ket{T_{a}}$ and $\ket{T_{b}}$. The gates $ X, Z$ and $H$ denote the Pauli and Hadamard gates respectively. $\bullet $ and $\oplus$ indicate the control and target qubits of $CNOT$ and $Toffoli$ gates respectively. All measurement outcomes ($ 0$ or $1$) are transmitted via classical channels represented by double lines into gates $X$ and $Z$. The numbers $1$, $2$, $3$ and $4$ with dashed red lines stand for the steps of performing the protocol [see Sec. \ref{sec}.]}
		\label{fig1}
	\end{center}
\end{figure}

In this section, we review briefly the proposed  protocol of Kiktenko $et$ $al.$ \cite{Kiktenko}, where the users initially  shared an entangled state of Bell type.
Let's consider that Alice and Bob share the Bell state :
\begin{equation}\label{bell}
\ket{B_{a,b}}=  \frac{1}{\sqrt2} (\ket{0_a0_b}+\ket{1_a1_b}),
\end{equation}
where $"a"$ and $"b"$ refer to Alice and Bob respectively. However, since we are interested in estimating  the teleported parameters by using  quantum Fisher information, we  have to describe the procedure of this protocol by means of the Bloch vectors. Therefore, we write the initial Bell state as:
\begin{equation}\label{eq:GW}
\rho_{B}=\frac{1}{4}(\hat{I}+\hat{\sigma_x}^{(a)}\hat{\tau_x}^{(b)}
-\hat{\sigma_y}^{(a)}\hat{\tau_y}^{(b)}+\hat{\sigma_z}^{(a)}\hat{\tau_z}^{(b)}),
\end{equation}
where $\hat{\sigma_i}^a$ and $\hat{\tau_i}^b$ (for $i=x,y,z$), represent the Pauli operators of Alice's qubit and Bob's qubit, respectively \cite{Metwally2009}.
\\
Let us assume that
Alice and Bob are given two  different states, $\ket{Q_a}$ and $\ket{Q_b}$  to be  bilateral teleported, such as
\begin{equation}\label{aq}
\ket{Q_a}=\cos(\frac{\theta_a}{2})\ket{0_a}+e^{i\phi_a}\sin(\frac{\theta_a}{2})\ket{1_a}, \quad
\ket{Q_b}=\cos(\frac{\theta_b}{2})\ket{0_b}+e^{i\phi_b}\sin(\frac{\theta_b}{2})\ket{1_b}.
\end{equation}
Similarly, these qubits can be represented by their Bloch vectors as:
\begin{eqnarray}{\label{qubits}}
\rho_q^{(a)}=\frac{1}{2}\Bigr(\hat{I}+\sum_{{i=x,y,z}}a_i\hat{\sigma_i}^{(a)}\Bigl), \quad \rho_q^{(b)}=\frac{1}{2}\Bigr(\hat{I}+\sum_{i=x,y,z}b_i\hat{\tau_i}^{(b)}\Bigl),
\end{eqnarray}
where $a_i$ and  $b_i$ (for $i=x,y,z$) are the Bloch vectors of Alice's qubit and Bob's qubit, respectively.
\\
The aim of the users is exchanging these states between them. Further, each  user has two different types of qubits; the trigger qubits which they are defined initially  as,
\begin{equation}
\ket{T_a,0}=\cos\Bigr(\frac{\tilde\theta_a}{2}\Bigl)\ket{0_a}+\sin\Bigr(\frac{\tilde\theta_a}{2}\Bigl)\ket{1_a}, \quad
\ket{T_b,0}=\cos\Bigr(\frac{\tilde\theta_b}{2}\Bigl)\ket{0_b}+\sin\Bigr(\frac{\tilde\theta_b}{2}\Bigl)\ket{1_b},
\end{equation}
these states become in the Bloch vectors like
\begin{equation}\label{trig}
\rho_{T_a}=\frac{1}{2}\Bigr(\hat{I}+\sum_{i=x,y,z} t_i^{(a)}\hat{\sigma_i}^{(a)} \Bigl),\quad
\rho_{T_b}=\frac{1}{2}\Bigr(\hat{I}+\sum_{i=x,y,z}t_i^{(b)}\hat{\tau_i}^{(b)}\Bigl),
\end{equation}
whereas $t_i^{(a)}$ $(t_i^{(b)})$ are the Bloch vectors of Alice's (Bob's) Trigger. Additionally, two storage qubits  for Alice, $\ket{S^{(1)}_a}, \ket{S^{(2)}_a}$ and two for Bob,  $\ket{S^{(1)}_b}, \ket{S^{(2)}_b}$. It is assumed that these qubits are initially prepared in a vacuum state, i.e., $\ket{S^{(i)}_j}=\ket{0}, i=1,2$ and $j=a,b$.
These states  may be represented  by Pauli operators as; $\rho_{s_a}^{(i)}=\frac{1}{2}(1+\hat{\sigma_z}^{(a)}), i=1,2$ and $\rho_{s_b}^{(i)}=\frac{1}{2}(1+\hat{\tau_z}^{(b)})$.
These types of qubits are used  as a storage of the projective measurement outcomes. All these types of qubits are described clearly in the circuit (Figure:$\ref{fig1}$).

\subsection{Local Operation}
The partners Alice and Bob perform two types of local measurements by the $CNOT$ gate and the Toffoli gate $(CCNOT)$. These two types of gates are used widely in the context of quantum purification \cite{Metwally2006}. It is more convenient to perform the $CNOT$ operations using Pauli's operators. However, in the computational basis, the $CNOT$ operation is defined as shown in Table (1).
In addition to the $CNOT$ operation, the users need to apply  what is called the Toffoli gate. This gate is defined such that the target qubit pairs change only when the two control qubits are
in the state $\ket{1}$. In a generic form, one can write its effect as
$CCNOT\ket{abc}=\ket{ab,c\oplus a.b}$, where  the $CCNOT$ gate may be defined by using the Pauli operators as,
\begin{eqnarray}\label{eq:CCN}
CCNOT&=&\frac{1}{4}\biggl[(1+\hat{\sigma}_z^{(1)})(1+\hat{\sigma}_z^{(2)})\hat{\sigma}_x^{(3)}+
(1+\hat{\sigma}_z^{(1)})(1-\hat{\sigma}_z^{(2)})
\nonumber\\
&&+
(1-\hat{\sigma}_z^{(1)})(1+\hat{\sigma}_z^{(2)})+
(1-\hat{\sigma}_z^{(1)})(1-\hat{\sigma}_z^{(2)})\biggr].
\end{eqnarray}

\begin{table}[h!]
 \begin{center}
\begin{tabular}{c|cccc}
\hline\hline
&$\hat{I}^{(2)}$&$\hat{\sigma}_x^{(2)}$&$\hat{\sigma}_y^{(2)}$&$\hat{\sigma}_z^{(2)}$\\
\hline
$\hat{I}^{(1)}$&$1$&$\hat{\sigma}_x^{(1)}\hat{\sigma}_x^{(2)}$&$\hat{\sigma}_y^{(1)}\hat{\sigma}_x^{(2)}$&$\hat{\sigma}_z^{(1)}$\\
$\hat{\sigma}_x^{(1)}$&$\hat{\sigma}_x^{(2)}$&$\hat{\sigma}_x^{(1)}$&$\hat{\sigma}_y^{(1)}$&$\hat{\sigma}_z^{(1)}\hat{\sigma}_x^{(2)}$\\
$\hat{\sigma}_y^{(1)}$&$\hat{\sigma}_z^{(1)}\hat{\sigma}_y^{(2)}$&$\hat{\sigma}_y^{(1)}\hat{\sigma}_z^{(2)}$&- $\hat{\sigma}_x^{(1)}\hat{\sigma}_z^{(2)}$&$\hat{\sigma}_y^{(2)}$\\
$\hat{\sigma}_z^{(1)}$&$\hat{\sigma}_z^{(1)}\hat{\sigma}_z^{(2)}$&$- \hat{\sigma}_y^{(1)}\hat{\sigma}_y^{(2)}$&
$\hat{\sigma}_x^{(1)}\hat{\sigma}_y^{(2)}$&$\hat{\sigma}_z^{(2)}$\\
 \hline
 \end{tabular}
 \vspace{0.1in}
 \caption{Bilateral $CNOT$ operation between the two qubits which is defined by
   $\hat{\sigma}_{\mu}^{(1)}$ and $\hat{\sigma}_{\mu}^{(2)}$.}
\end{center}
\end{table}

\subsection{Performing the Protocol}\label{sec}
In the following, we describe the duty of the users to perform the suggested protocol step by step as shown in Figure \ref{fig1}.

\begin{itemize}
\item{\it From Alice to Bob}

\begin{enumerate}
	\item {\it Step one}:\\ The users prepare their initial state, which consists of 10 qubits, by using the states Eqs.(\ref{bell}-\ref{trig}). This state may be written as:
\begin{equation}
\ket{\psi_{s}}=\ket{B_aQ_a S^{(1)}_a S^{(2)}_aT_a} \ket{B_bQ_b S^{(1)}_b S^{(2)}_bT_b}.
\end{equation}

\item {\it Step two:}\\The users apply the $CNOT$ gate with $\ket{Q_{a,b}}$ are control qubits and the Bell qubits $\ket{B_{a,b}}$ as target qubits as,
\begin{eqnarray*}
CNOT\ket{\psi_s}=CNOT\ket{B_aQ_a S^{(1)}_a S^{(2)}_aT_a}CNOT \ket{B_bQ_b S^{(1)}_b S^{(2)}_bT_b},
\end{eqnarray*}
followed by  applying  the Hadamard gates on the qubits $\ket{Q_{a,b}}$ in the previous output results
\begin{eqnarray}
H\Bigr(CNOT\ket{\psi_s}\Bigl)=H\Bigr(CNOT\ket{B_aQ_a S^{(1)}_a S^{(2)}_aT_a}\Bigl)H\Bigr(CNOT \ket{B_bQ_b S^{(1)}_b S^{(2)}_bT_b}\Bigl).
\end{eqnarray}

\item{Step three:\\}
 Alice and Bob perform two $CCNOT$ gate. In the first one, the triggers $\ket{T_{a,b}}$ and the Bell pairs $\ket{B_{a,b}}$ play the role of the control qubits and $\ket{ S_{a,b}^{(2)}}$ as a target qubit. While in the second $CCNOT$ the triggers $\ket{T_{a,b}}$ and $\ket {Q_{a,b}}$ as a control qubits and $\ket{ S_{a,b}^{(1)}}$ is the target qubit
to get:
\begin{eqnarray}
CCNOT\Bigl(H\Bigr(CNOT\ket{\psi_s}\Bigl)\Bigr)=CCNOT\Bigl(H\Bigr(CNOT\ket{B_aQ_a S^{(1)}_a S^{(2)}_aT_a}\Bigl)\Bigl)\otimes
\nonumber\\
CCNOT\Bigr(H\Bigr(CNOT \ket{B_bQ_b S^{(1)}_b S^{(2)}_bT_b}\Bigl)\Bigl)
\end{eqnarray}

\item{Step Four:\\}  The measurement's results are stored in the qubits $\ket{S_a^{(1)}}$  and  $\ket{S_a^{(2)}}$. These results  are sent by classical channel  to Bob. Based on Alice's measurements, Bob performs the  adequate local operations ($\hat{X} ,\hat{Y},\hat{ Z}, \hat{I}$), on his Bell state particle to recover  the state $\ket{Q_a}$.
\end{enumerate}
\item{\it From Bob to Alice:\\}
Bob may  perform the same operation that Alice has done to teleport his state $\ket{Q_b}$ to Alice.
\end{itemize}
\subsection{Implementation of the circuit}
The final output of the circuit depends on the initial state of the triggers. However, we have the following possibilities:
\begin{enumerate}
\item The initial state of both triggers are initially prepared in the states, $\rho_{T_a}=\frac{1}{2}\Bigl(1+\hat{\sigma_z}^{(a)}\Bigr)$ and $\rho_{T_b}=\frac{1}{2}\Bigl(1+\hat{\tau_z}^{(b)}\Bigr)$. In this case, each partner obtains the completely mixed state, i.e., $\rho^{(a)}_q=\rho^{(b)}_q=\rho_{0}=\frac{1}{2}(\ket{0}\bra{0}+\ket{1}\bra{1})$
\item  Alice's and Bob's triggers are prepared in the states  $\rho_{T_a}=\frac{1}{2}\Bigl(1-\hat{\sigma_z}^{(a)}\Bigr)$ and $\rho_{T_b}=\frac{1}{2}\Bigl(1-\hat{\tau_z}^{(b)}\Bigr)$.
In this case, Alice and Bob perform a perfect projective measurement  with probability:
\begin{equation}
p_i=tr\Bigl\{\rho_{T_a}\rho_q^{(a)}\Bigr\}=tr\Bigl\{\rho_{T_b}\rho_q^{(b)}\Bigr\}=\sin^2(\theta_i/2),\quad\quad~~ i=a,b.
\end{equation}
The users will end up their protocol by two states, one of them at Alice's side.
\begin{eqnarray}\label{tele1}
\rho^{(a)}_{tel}=p_b\bar p_a\rho_q^{(b)}+(1-p_b\bar p_a)\rho_{0},~~ \bar p_a=1-p_a
\end{eqnarray}
and the other in Bob's hand
 \begin{eqnarray}\label{tele2}
\rho^{(b)}_{tel}=p_a\bar p_b\rho_q^{(a)}+(1-p_a\bar p_b)\rho_{0},~~ \bar p_b=1-p_b
\end{eqnarray}

\item If Alice's trigger qubit is prepared in  the state  $\rho_{T_a}=\frac{1}{2}\Bigl(\hat{I}-\hat{\sigma_z}^{(a)}\Bigr)$ and Bob's  trigger qubit is prepared in the state
 $\rho_{T_b}=\frac{1}{2}\Bigl(\hat{I}+\hat{\tau_z}^{(b)}\Bigr)$, then only Alice  has the ability to teleport her qubit to Bob. While, Alice get the completely mixed state i.e., the final state at Alice's hand is $\rho_{tel}^{(a)}=\rho_{0}$ and at Bob's hand $\rho_{tel}^{(b)}=\rho_q^{(a)}$.
\end{enumerate}
\subsection{Fidelity of teleportation}\label{section2.5}
To quantify the efficiency of the suggested teleportation protocol, we quantify the  fidelity $\mathcal{F}$ of the teleported state. It
measures the similarity between the input state $\rho_{q}^{(a)}$ ($\rho_{q}^{(b)}$) and the output state $\rho_{tel}^{(b)}$ ($\rho_{tel}^{(a)}$) respectively. The expression of the fidelity $\mathcal{F}$ is defined as $\cite{Wang,allati}$ :
\begin{equation}
\mathcal{F}= <Q^{i}| \rho_{out} |Q^{i}>, \quad  i=a,b
\end{equation}
with $|Q^{i}>$ and $\rho_{out}$ are the input and the output states respectively. In general, the input states are unknown. The averaged fidelity over all the input states may be described by
\begin{equation}
\mathcal{F}_{avg}=\frac{1}{4\pi}\int_{0}^{\pi}d\theta_i\int_{0}^{2\pi}\mathcal{F}(\theta_i,\phi_i,p_a,p_b)\sin\theta_i d\phi_i~, \quad i=a,b
\end{equation}
where:
\begin{equation}
\mathcal{F}(\theta_i,\phi_i,p_a,p_b)=|c|^2_i\Bigl(A_2+A_1|c|^2_t\Bigr)+A_1\Bigl(c^*_is_ic_ts^*_t+s^*_ic_is_tc^*_
t\Bigr)+|s|^2_i\Bigl(A_1|s|^2_t+A_2\Bigr),
\end{equation}
with different elements $c_i=\cos(\frac{\theta_i}{2}),~s_i=\sin(\frac{\theta_i}{2})e^{i\phi_i}, c_t=\cos(\frac{\theta_t}{2})$ and $s_t=\sin(\frac{\theta_t}{2})e^{i\phi_t}$. Whereas, $i$ and $t$ indicate for the initial and the teleported state respectively. The terms $A_1$ and $A_2$ depend on the direction of teleportation. i.e, from Alice to Bob $A_1=p_a\bar{p_b}$, $A_2=\frac{1-p_a\bar{ p_b}}{2}$. While, from Bob to Alice $A_1=p_b\bar{p_a}$ and $ A_2=\frac{1-p_b\bar{ p_a}}{2}$.
\\[1 cm]
The fidelity of teleportation  between the two users, $\mathcal{F}^{A\to B} $ and  $\mathcal{F}^{B\to A} $ are given by:
\begin{equation}\label{Fidelities}
\mathcal{F}^{(A\to B)}= \frac{1}{2}(1+p_a\bar{p_b}),\quad \mathcal{F}^{(B\to A)}= \frac{1}{2}(1+p_b\bar{p_a}),\quad
\end{equation}
where the probabilities  may be written as:
\begin{eqnarray}\label{probabilities}
    p_a&=&\frac{1}{2}\Bigr(1+\cos({\tilde\theta_a})\cos({\theta_{a}})+\sin(\tilde\theta_a)\sin(\theta_a)\cos{\phi_{a}}+\sin({\tilde\theta_a})\sin({\theta_{a}})\sin{\phi_{a}}\Bigl),
\nonumber\\
    p_b&=&\frac{1}{2}\Bigr(1+\cos({\tilde\theta_b})\cos({\theta_{b}})+\sin(\tilde\theta_b)\sin(\theta_b)\cos{\phi_{b}}+\sin({\tilde\theta_b})\sin({\theta_{b}})\sin{\phi_{b}}\Bigl).
\end{eqnarray}
Now, by using Eqs.(\ref{Fidelities}) and (\ref{probabilities}), one can obtain the Fidelity of the teleported state for both directions at an initial state setting.\\

\begin{figure}[!htb]
\begin{center}
\includegraphics[scale=.65]{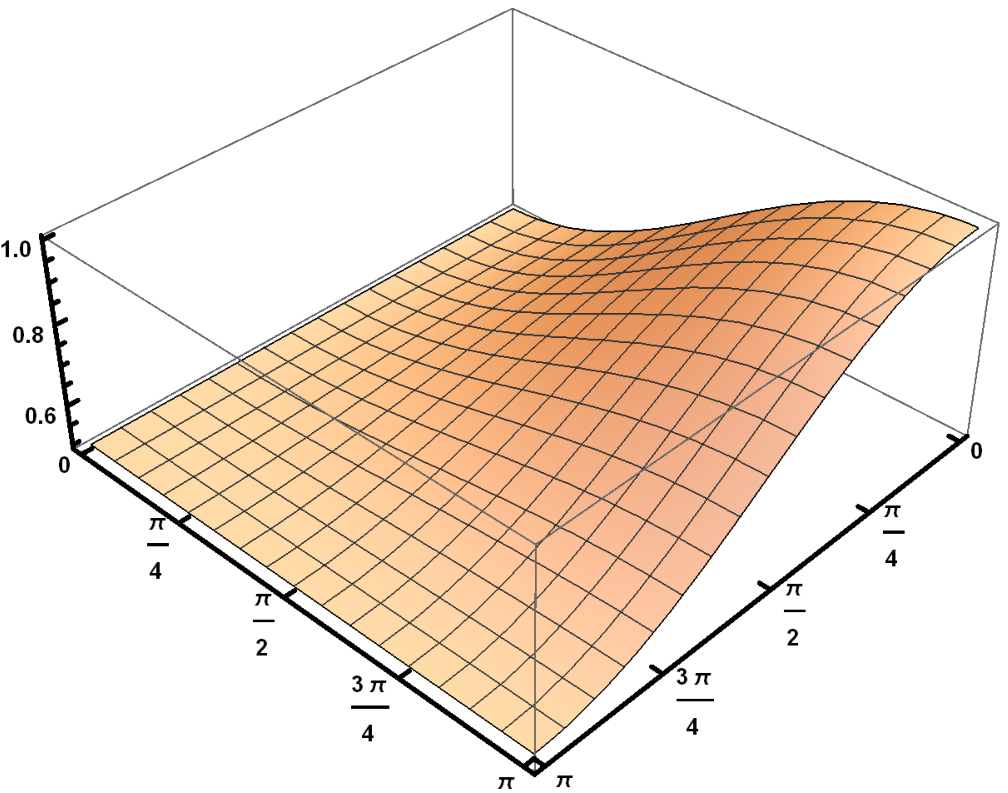}
\put(-220,80){$\mathcal{F}^{(A\to B)}$}
                      \put(-100,160){$(a)$}
                      \put(-160,25){$\tilde\theta_{b}$}
                      \put(-35,20){$\tilde\theta_{a}$}~\quad\quad\quad\quad
                      \includegraphics[scale=.45]{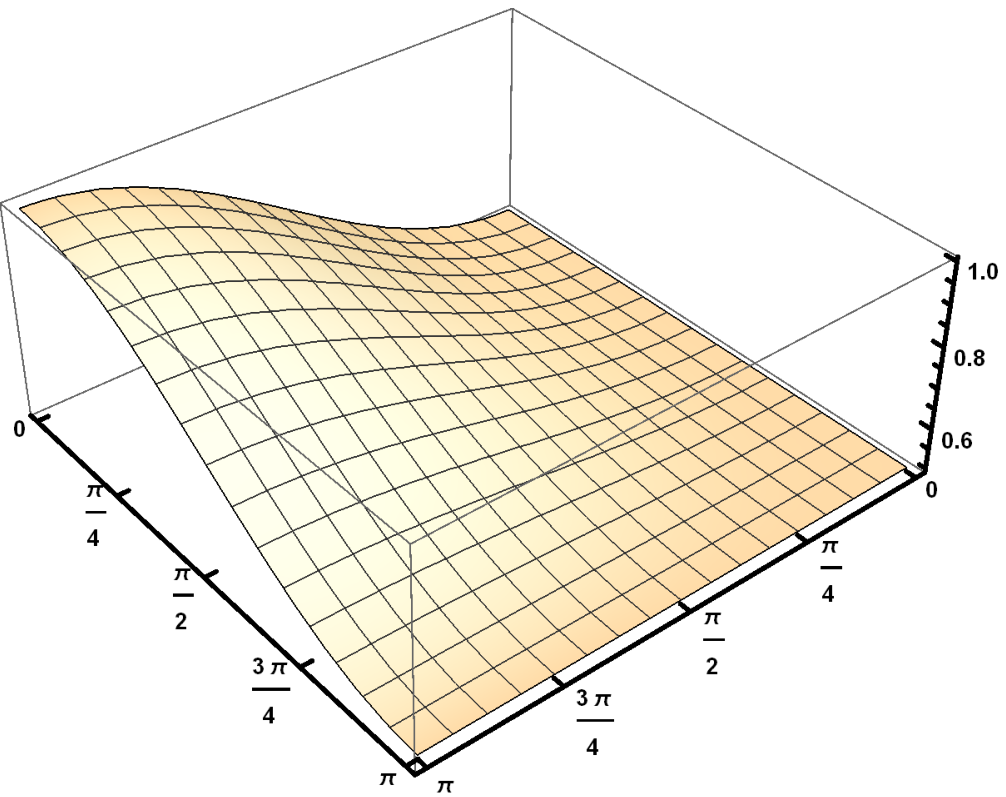}
\put(-3,77){$\mathcal{F}^{(B\to A)}$}
                      \put(-100,160){$(b)$}
                      \put(-165,30){$\tilde\theta_{b}$}
                      \put(-45,20){$\tilde\theta_{a}$}
\caption{Bidirectional teleportation Fidelity, where the initial states are prepared such that $\theta_a = \theta_b = 0$ of the initial qubits $Q_a$ and $Q_b$. (a) The fidelity of teleportation $\mathcal{F}^{(A\to B)}$ and (b) represents $\mathcal{F}^{(B\to A)}$. }
\label{Fig1}
 \end{center}
\end{figure}

Fig. (2), shows the behavior of the Fidelity of the teleported state from both directions, namely {$\mathcal{F}^{(A\to B)}$} and {$\mathcal{F}^{(B\to A)}$}, where it is assumed that the teleported state is initially prepared in the state $\rho^{(a)}_{q}=\frac{1}{2}(\hat{I}+\hat\sigma_z^{(a)})$ at Alice side, while Bob's teleported state is given by $\rho^{(b)}_{q}=\frac{1}{2}(\hat{I}+\hat\tau_z^{(b)})$, namely  the users teleport  only classical information, i.e. $\theta_a=\theta_b=0$, while the triggers are prepared in an arbitrary state with $\phi=0$. It is clear that,  from Fig.(2a), at any value of $\tilde\theta_b$,  the Fidelity $\mathcal{F}^{(A\to B)}$ increases gradually as $\tilde\theta_a$ increases. The maximum Fidelity is reached at $\tilde\theta_a=\pi$, namely the  Alice's trigger is prepared in the state $\rho_{T_b}=\frac{1}{2}(\hat{I}-\hat\tau_z^{(b)})$. However, {$\mathcal{F}^{(A\to B)}$}, decreases gradually as $\tilde\theta_b$ increases. The minimum values of {$\mathcal{F}^{(A\to B)}$} are  displayed at $\tilde\theta_a=\pi$. Fig.(2b), displays the behavior of the fidelity from Bob to Alice, {$\mathcal{F}^{(B\to A)}$}, where the teleported state is prepared in the state $\rho^{(b)}_{q}=\frac{1}{2}(\hat{I}+\hat\tau_z^{(b)})$. The behavior of  $\mathcal{F}^{(A\to B)}$ is similar to that predicted in Fig.(2a), where the increasing and decreasing of the fidelity depends on the state of Bob's trigger, namely the maximum value is predicted at $\rho_{T_b}=\frac{1}{2}(\hat{I}-\hat\tau_z^{(b)})$.

\begin{figure}[!htb]
\begin{center}
\includegraphics[scale=.65]{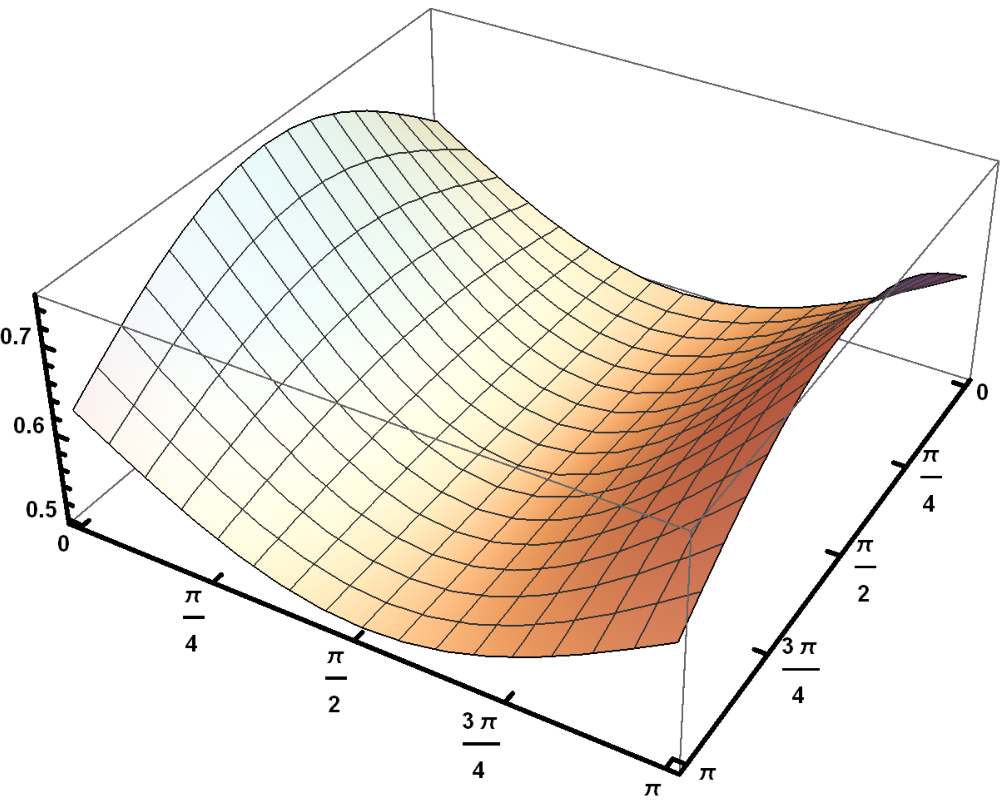}
\put(-220,70){$\mathcal{F}^{(A\to B)}$}\put(-100,160){$(a)$}
                      \put(-140,20){$\tilde\theta_{b}$}
                      \put(-20,40){$\tilde\theta_{a}$}~\quad\quad\quad\quad
                      \includegraphics[scale=.50]{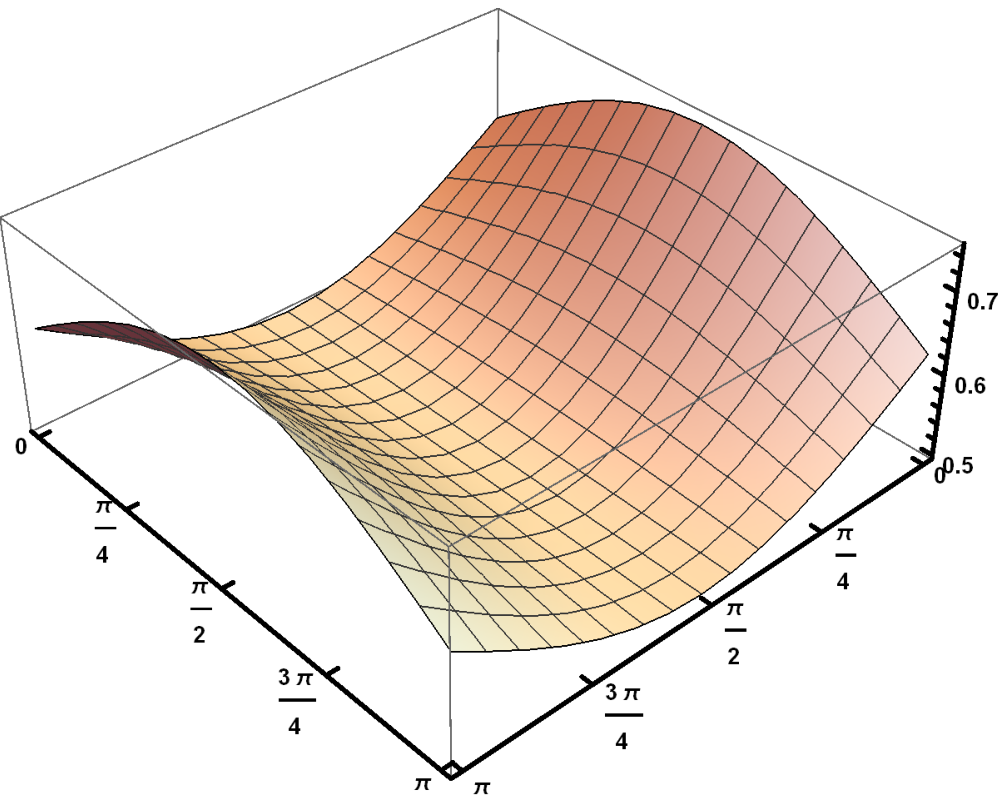}
\put(-3,85){$\mathcal{F}^{(B\to A)}$}\put(-100,170){$(b)$}
                      \put(-180,30){$\tilde\theta_{b}$}
                      \put(-50,25){$\tilde\theta_{a}$}
\caption{The same as Fig.(2), but the initial teleported state is prepared in the state $\ket{\psi_q^a}=\ket{\psi_q^b}=\frac{1}{\sqrt{2}}(\ket{0}+\ket{1})$, namely $\theta_a=\theta_b=\pi/2$.}
 \end{center}
\end{figure}

Teleporting of  quantum information  via the bidirectional protocol is explored  in Fig.(3), where it is assumed that Alice and Bob prepared their states as $\ket{\psi_q^a}=\ket{\psi_q^b}=\frac{1}{\sqrt{2}}(\ket{0}+\ket{1})$, namely we set $\theta_a=\theta_b=\pi/2$.  It is clear that the maximum values of $\mathcal{F}^{(A\to B)}$ and $\mathcal{F}^{(B\to A)}$ are smaller than those shown in Fig.(2). The behavior of both Fidelities is similar, where they  decrease gradually  as the weight of both triggers increases. The minimum values of $\mathcal{F}^{(A\to B)}$ and  are observed at $\tilde\theta_b=\pi/2$, while  it is increased gradually when Alice's trigger angle $\tilde\theta_a$  increases.  The state of both triggers have the same effect on the behavior of $\mathcal{F}^{(B\to A)}$.\\

\begin{figure}[!htb]
\begin{center}
\includegraphics[scale=.70]{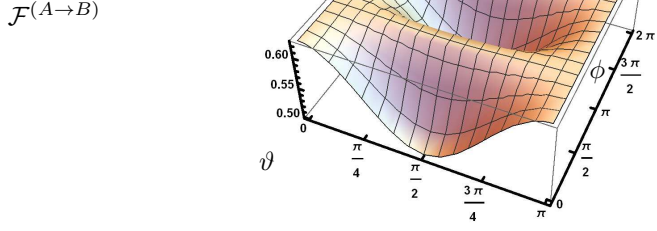}
\put(-255,75){$\mathcal{F}^{(A\to B)}$}
                      \put(-160,20){$\vartheta$}
                      \put(-35,55){$\phi$}

\caption{ Fidelity of teleportation between Alice and Bob for ($\phi_a = \phi_b = \phi)$ of the initial qubits $Qa$ and $Qb$ and $\tilde\theta_a = \tilde\theta_b = \tilde\theta_t=\vartheta$. the Fidelity of teleportation  reaches the value of 0.62 as a maximal value.}
\label{Fig3}
              \end{center}

\end{figure}

In Fig.(4a), we investigate the behavior of $\mathcal{F}^{(A\to B)}$, where we assume that the teleported qubit, Alice and Bob's triggers are prepared on the same state, namely $\theta_a=\tilde\theta_a=\tilde\theta_b=\tilde\theta_t=\vartheta$. The behavior of the fidelity shows that, the maximum bound of $\mathcal{F}^{(A\to B)}$ is smaller than that displayed in Fig.(2a). The Fidelity decreases gradually as the weight angles of all the qubit decreases. The minimum bounds are predicted at $\vartheta=\frac{\pi}{2}$, then increases gradually  to reach their maximum values at $\vartheta=\pi$.
From this figure, it is clear that, the possibility of applying the bidirectional teleportation protocol to teleport classical data is much better than teleporting quantum data. This is predicted at $\vartheta=\frac{\pi}{2}$, namely the initial state of the Alice' qubit is prepared in the state, $\ket{\psi_q^a}=\frac{1}{\sqrt{2}}(\ket{0}+e^{i\phi}\ket{1})$.  Moreover, as the phase $\phi$ increases, the Fidelity increase gradually  to reach its maximum value at $\phi\simeq\frac{3\pi}{4}$ at any value of the $\vartheta$. In further values of the phase $\phi$, the Fidelity decreases gradually to reach its minimum values at $\phi\simeq\frac{5\pi}{4}$. This behavior is periodically repeated.

\begin{figure}[!htb]
\begin{center}
\includegraphics[scale=.65]{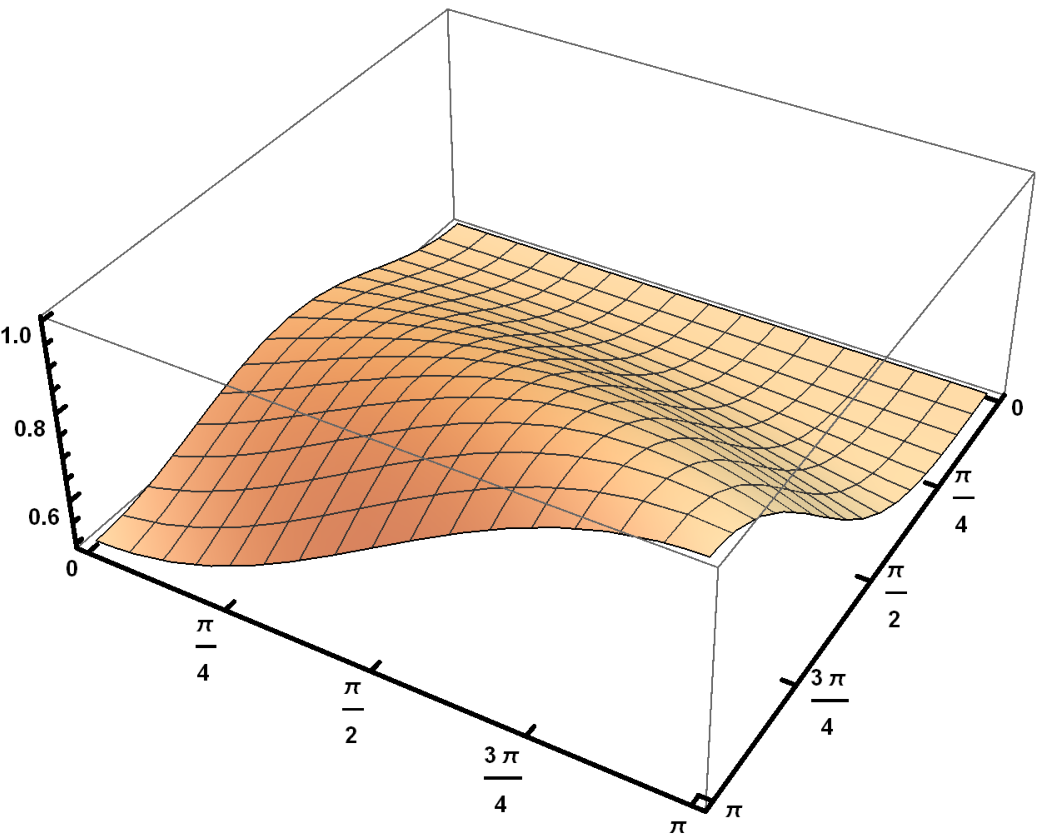}
\put(-230,70){$\mathcal{F}^{(A\to B)}$}\put(-100,160){$(a)$}
                       \put(-140,15) {$\theta_{a}$}
                      \put(-20,30){$\tilde\theta_{t}$}~~\quad\quad\quad\quad
                     \includegraphics[scale=.65]{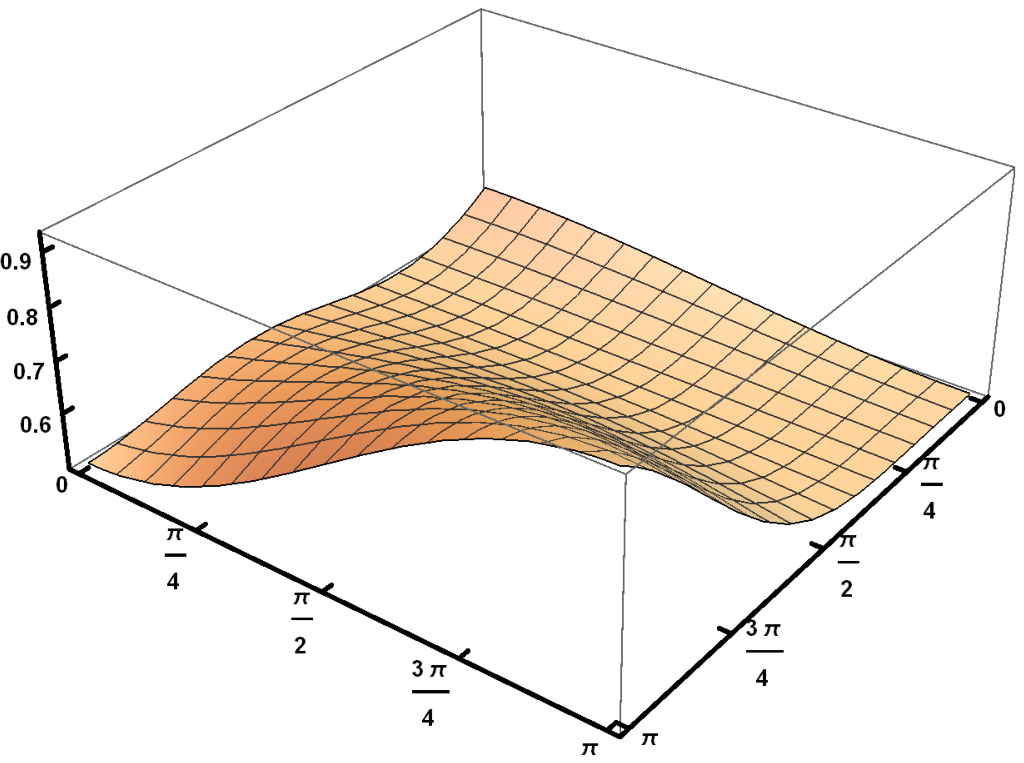}
                     \put(-225,75){$\mathcal{F}^{(A\to B)}$}\put(-100,160){$(b)$}
                    \put(-140,15){$\theta_{a}$}
                      \put(-20,30){$\tilde\theta_{t}$}
\caption{ The Fidelity, $\mathcal{F}^{(A\to B)}(\theta_a,\tilde\theta_t)$, (a):~$\theta_b=0$ and (b): $\theta_b=\pi/4$.}
\label{Fig3}
              \end{center}

\end{figure}

In the previous investigation, it is assumed that both initial qubits are prepared in the same state. In what follows, we discuss the possibility of bidirectional teleportation, when the partners have different initial states. For this aim, we consider the following figure, where it is assumed that Bob's qubit is initially prepared  in the state $\rho_b=\ket{0}\bra{0}$, i.e., $\theta_b=\pi$, while Alice's qubit and her triggers are prepared in an arbitrary state with $\phi_a=\tilde\phi=0$.
Fig.(5a), shows the behavior of the Fidelity $\mathcal{F}^{(A\to B)}(\theta_a,\tilde\theta_t)$, where the teleported Bob's qubit is prepared in the state $\rho_q^{(b)}=\frac{1}{2}(\hat{I}-\hat\tau_z^{(b)})$, i.e $\theta_b=\frac{\pi}{2}$. The Fidelity's behavior shows that, $\mathcal{F}^{(A\to B)}$ increases as the weight angles of the teleported state and Alice' trigger  increase.  Therefore, the maximum Fidelity is predicted at $\theta_a=\tilde\theta_t=\pi$, namely $\rho_q^{(a)}=\frac{1}{2}(\hat{I}-\hat\sigma_z)$. The effect of the initial state of the trigger shows that at any initial state settings of Alice's qubit, the $\mathcal{F}^{(A\to B)}$ increases to reach its maximum bounds. However, at further values of $\tilde\theta_t,$, the Fidelity decreases gradually, where the minimum values are predicted at $\tilde\theta_t=0$, i.e., the trigger is prepared in the state $\rho_{T_a}=\frac{1}{2}(\hat{I}+\hat\sigma_z^{(a)})$.
In Fig.(5b), we investigate the effect of different initial state settings of  Bob's state, where it is assumed that $\rho_q^{(b)}$ is prepared at $\theta_b=\frac{\pi}{4}$. The  behavior is similar to that predicted in Fig.(5a), but the maximum bounds are smaller than those displayed in Fig.(5a).  Moreover, the decay rate is larger compared with that displayed at $\theta_b=0$.
In this context, we would like to mention that, the same behavior is predicted for $\mathcal{F}^{(B\to A)}(\theta_b, \tilde\theta_t)$.\\

From Fig(5), one may conclude that the bidirectional Fidelity from  Alice to Bob and vice versa, depends on the initial state settings of all the three qubits. However, the maximum Fidelity is attended when all the qubits polarized in the same direction. Moreover, the Fidelity of teleporting classical information is much better than teleporting quantum information.

\section{Quantum Fisher Information (QFI) and Quantum Fisher information matrix (QFIM)}
Many quantum parameters cannot be measured directly, but quantum Fisher information (QFI) may be used to estimate these parameters.  Moreover, in the context of  quantum teleportation  process, it is  shown that by teleporting the relevant parameters encoded in the teleported state, one could measure the credibility of the protocol by measuring QFI \cite{El Anouz}

\subsection{Single parameter estimation }
Many quantum parameters cannot be measured directly, but quantum Fisher information(QFI) may be used to estimate these parameters.  Moreover, in the context of  quantum teleportation  process, it is  shown that by teleporting the relevant parameters encoded in the teleported state, one could measure the credibility of the protocol by measuring QFI \cite{El Anouz}.\\

To calculate the QFI one can use the Bruse distance and the Uhlmann fidelity \cite{21} or using the spectral decomposition of the state $\rho$. A simple expression is given in \cite{Xiao} for any single qubit state $\rho=\frac {1}{2}(1+\overrightarrow{s}.\hat{\sigma})$. Whereas $\overrightarrow{s}=(s_{x}, s_{y}, s_{z})$ describes  the real Bloch vector and $\hat{\sigma}=(\hat{\sigma}_{x}, \hat{\sigma}_{y}, \hat{\sigma}_{z} ) $ denotes the
Pauli matrices.
\\[1cm]
The amount Fisher information  $\mathcal{F_{\theta}}$ with respect to the parameter $\theta$ is written as follows:

\begin{equation}\label{BlochQF}
\mathcal{F}_{\theta} = \left\{
\begin{array}{ll}
|\partial_{\theta} \overrightarrow{s}|^{2}+\frac{(\overrightarrow{s}.\partial_{\theta}\overrightarrow{s})^{2}}{1-|\overrightarrow{s}|^{2}},  & \mbox{if } |\overrightarrow{s}|<1 \\
|\partial_{\theta} \overrightarrow{s}|^{2}, & \mbox{if }
|\overrightarrow{s}|=1,
\end{array}
\right.
\end{equation}
with: $\partial_{\theta}=\partial/\partial_{\theta}$. From Eq. (\ref{BlochQF}) the quantum Fisher information of a pure state ($ |\overrightarrow{s}|=1$) is given by the expression $|\partial_{\theta} \overrightarrow{s}|^{2}$. While, for a mixed state ($|\overrightarrow{s}|<1$)
is presented by $|\partial_{\theta} \overrightarrow{s}|^{2}+\frac{(\overrightarrow{s}.\partial_{\theta}\overrightarrow{s})^{2}}{1-|\overrightarrow{s}|^{2}}$.
\\
According to Ref. \cite{Liu},  the QFI of a single qubit state $\ket{\psi}$ is defined as:
\begin{equation}\label{QFI1}
\mathcal{F}_{\theta} = 4 ( <\partial_{\theta} \psi |\partial_{\theta} \psi>  - |<\partial_{\theta} \psi|\psi>|^2 ),
\end{equation}
the expression of the quantum Fisher information in Eq.(\ref{QFI1}) can be obtained using only the term of the pure state $\ket{\psi}$. The QFI in the two previous equations (\ref{BlochQF}) and (\ref{QFI1}) provides the estimation precision limit of only one parameter, through the $Cramer-Rao$ inequality $(Var(\hat\theta)  \geq  \mathcal{F}^{-1})$ \cite{paris}; with $Var(\hat\theta)$ is the variance of the parameter and $\mathcal{F}$ is the quantum Fisher information.
where the larger QFI represents the better estimation precision. However, the inverse of the QFI for one unknown parameter provides the lower error limit of the parameter estimation . Our main goal is to reach the smallest value of the variance of the parameter.\\

Now, let us assume that Alice and Bob have prepared their qubits in the states Eq.(\ref{qubits}), while the triggers qubits are  defined  by Eq.(\ref{trig}).  The users use the bidirectional protocol as described in (Sec.\ref{section2}), where the final teleported states between the two users are given by the states (\ref{tele1}) and (\ref{tele2}).  In this context, the final teleported states depend on the initial states of all the used qubits. Therefore, our main aim of this section is to investigate the effect of these parameters on the initial weight angle of the teleported state from Alice to Bob or vice versa.   However, by using the definition of the quantum Fisher information for a single qubit, (\ref{BlochQF}) and (\ref{QFI1}), the Fisher information with respect to the weight angle of  Alice and Bob qubits are given by $\mathcal{F}_{\theta_k}=1,  k=a(b)$. However, the quantum Fisher information for the phase parameter of both qubits is defined as, $\mathcal{F}_{\phi_k}=\sin(\theta_k)^2, k=a(b)$.
In Bloch representation form, the final state at Alice's and Bob's hands are described by the following Bloch vectors.

\begin{equation}\label{10}
a_{x}^{(tel)}=2p_a \bar p_b\cos\phi_a \cos\Bigr(\frac{\theta_a}{2}\Bigl) \sin\Bigr(\frac{\theta_a}{2}\Bigl)\quad,\quad
a_{y}^{(tel)}= - 2p_a \bar p_b\sin\phi_a \cos\Bigr(\frac{\theta_a}{2}\Bigl) \sin\Bigr(\frac{\theta_a}{2}\Bigl)\quad,\quad
a_{z}^{(tel)}=p_a \bar p_b\cos(\theta_a),
\end{equation}
Alice received the state $(b_{x}^{(tel)},b_{y}^{(tel)},b_{z}^{(tel)})$ by replacing $p_a \bar p_b\leftrightarrow p_b \bar p_a$ and $\theta_a \leftrightarrow \theta_b$, $\phi_a \leftrightarrow \phi_b$ in Eq. (\ref{10}):
\begin{equation}\label{8}
b_{x}^{(tel)}=2p_b \bar p_a\cos\phi_b \cos\Bigr(\frac{\theta_b}{2}\Bigl) \sin\Bigr(\frac{\theta_b}{2}\Bigl)\quad,\quad
b_{y}^{(tel)}=-2p_b \bar p_a\sin\phi_b \cos\Bigr(\frac{\theta_b}{2}\Bigl) \sin\Bigr(\frac{\theta_b}{2}\Bigl)\quad,\quad
b_{z}^{(tel)}=p_b \bar p_a\cos(\theta_b),
\end{equation}

\begin{figure}[!htb]
\begin{center}
\includegraphics[scale=.60]{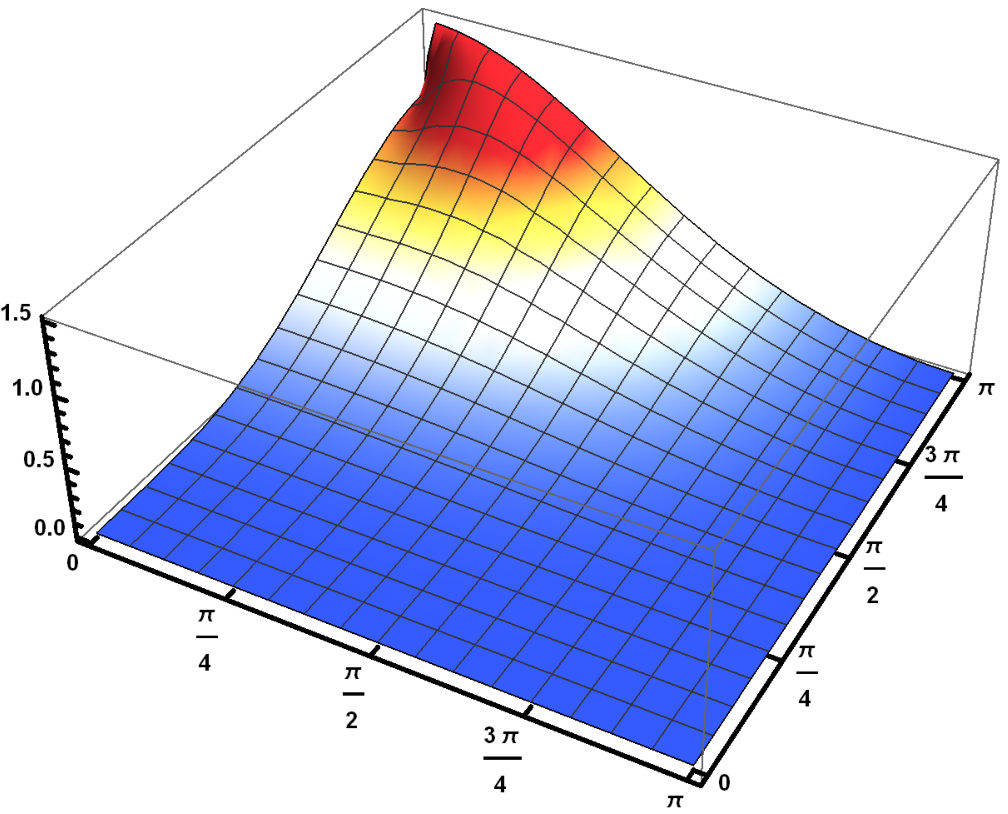}
\put(-120,150){$(a)$}
\put(-190,70){$\mathcal{F_{\theta}}_a$}
                      \put(-130,20){$\theta_{a}$}
                      \put(-20,30){$\theta_{b}$}~\quad\quad\quad
\includegraphics[scale=.60]{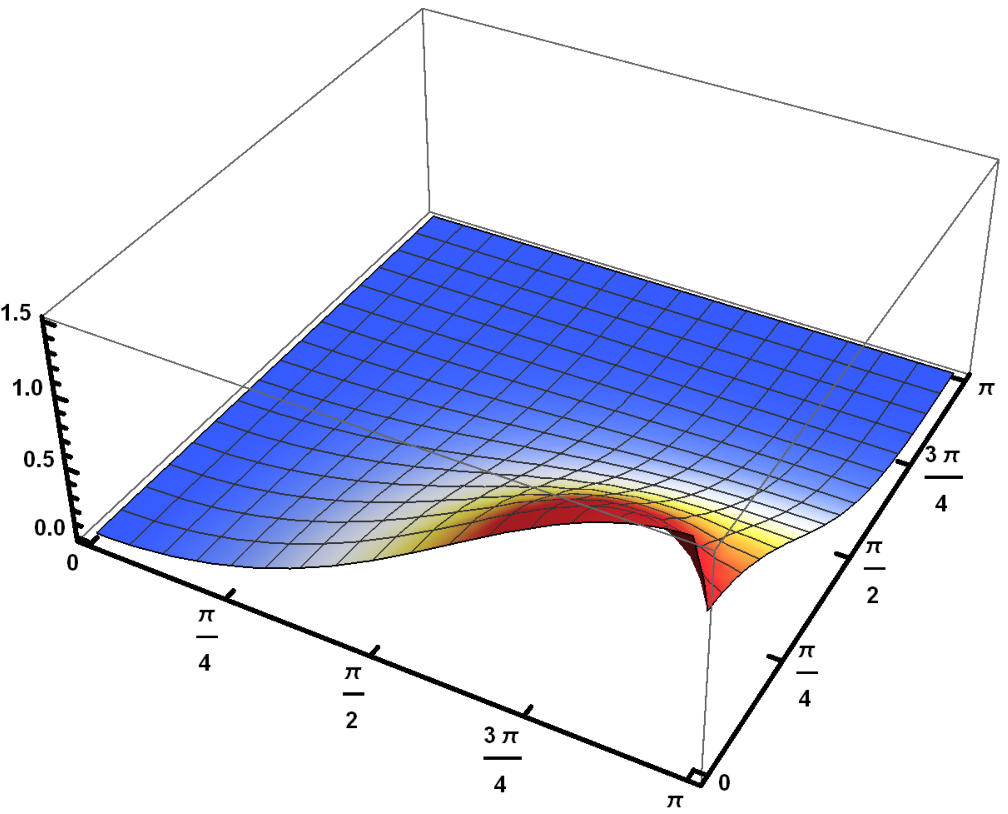}\put(-120,150){$(b)$}
\put(-190,70){$\mathcal{F_{\theta}}_a$}
                      \put(-130,20){$\theta_{a}$}
                      \put(-20,30){$\theta_{b}$}~\quad\quad\quad
\includegraphics[scale=.50]{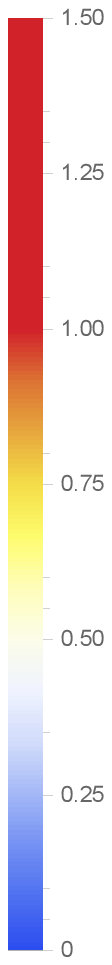}

\caption{Fisher information of $\theta_a$, with arbitrary phase where  (a):  at  $\tilde\theta_a=\tilde\theta_b=0$ and   (b): at  $\tilde\theta_a=\tilde\theta_b= \pi$ . }
\label{Figure4}
              \end{center}
\end{figure}
In Fig.(6), we display the behavior of the teleported Fisher information with respect to the weight parameter $\theta_a$, $\mathcal{F}_{\theta_a}(\theta_a,\theta_b)$. Where we consider that the triggers qubits are polarized either on the state
$\rho_T=\frac{1}{2}(\hat{I}+\hat\sigma_z)$ or $\rho_T=\frac{1}{2}(\hat{I}-\hat\sigma_z)$.  This figure illustrates that, the maximum bounds of the quantum Fisher information, $\mathcal{F}_{\theta_a}$  are predicted when both qubit and its  trigger  have the same polarization direction. As it is displayed from Fig.(6a), the maximum values are predicted at $\theta_a=0$, namely  $\rho_q^{(a)}=\frac{1}{2}(\hat{I}+\hat\sigma^{(a)}_{z})$ and decreases gradually as $\theta_a$ increases.  On the other hand, the initial state settings of Bob state play an important role in the assessment of  maximum values of $\mathcal{F}_{\theta}$.  However, if Bobs's qubit and Alice' trigger have the same polarization one may estimate the initial weight parameter of $\theta_a$ with high efficiency. The same behavior is displayed in Fig.(6b), where it is assumed that Alice's and Bob's triggers are prepared in the state $\rho_T=\frac{1}{2}(\hat{I}-\hat\tau_z), t=a,b$ . In this case the $\mathcal{F}_{\theta_a}$ increases as $\theta_a$ increases while $\theta_b$ decreases and the maximum bounds are reached if  Alice's qubit and its trigger are prepared in the same state.

\begin{figure}[!htb]
\begin{center}
\includegraphics[scale=.60]{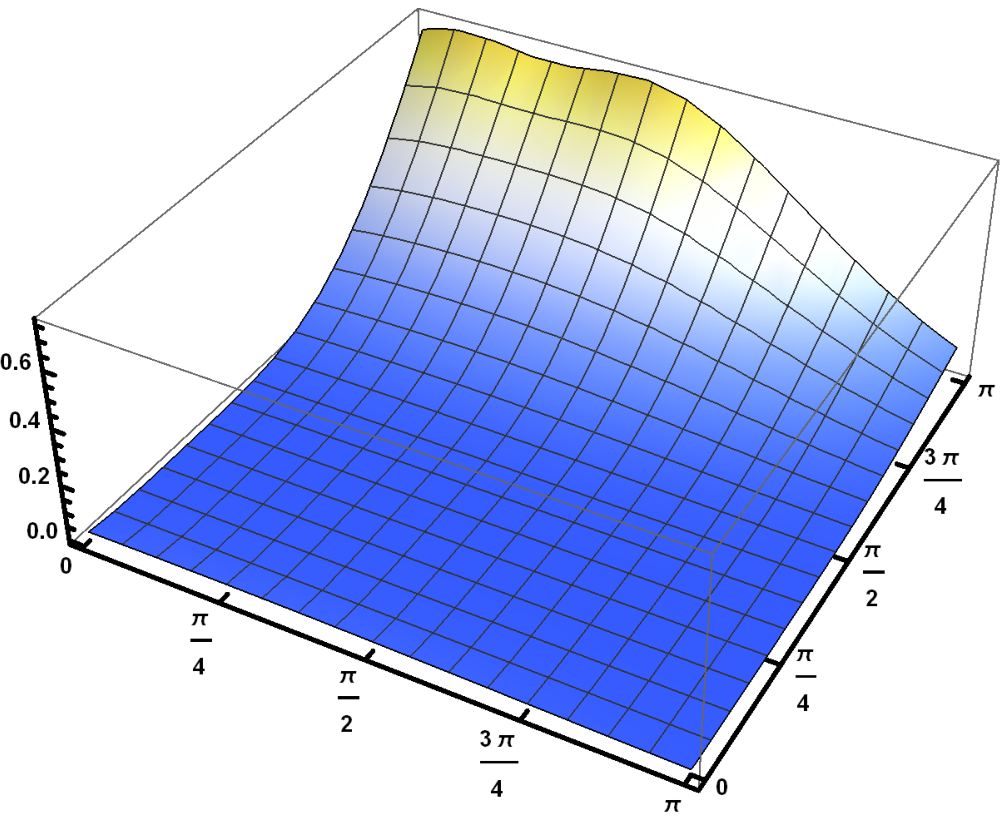}
\put(-120,150){$(a)$}
\put(-190,90){$\mathcal{F_{\theta}}_a$}
                      \put(-130,20){$\theta_{a}$}
                      \put(-20,30){$\theta_{b}$}~\quad\quad\quad
\includegraphics[scale=.60]{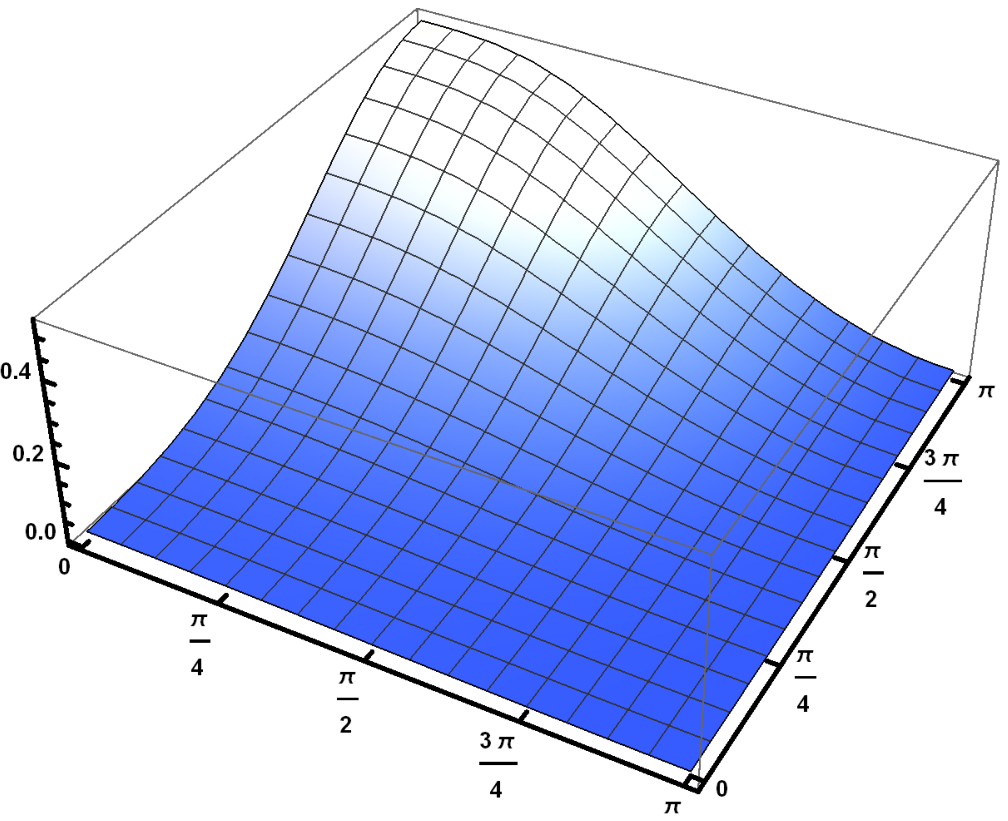}\put(-120,150){$(b)$}
\put(-190,90){$\mathcal{F_{\theta}}_a$}
                      \put(-130,20){$\theta_{a}$}
                      \put(-20,30){$\theta_{b}$}~\quad\quad\quad
\includegraphics[scale=.50]{legend.eps}
\caption{Fisher information $\mathcal{F}_{\theta_a}$, with  $\tilde\theta_a=\tilde\theta_b=\frac{\pi}{4}$  where we set  (a): $\phi = 0$.  (b): $\phi = \frac{\pi}{2}$ }

\label{Figure5}
              \end{center}
\end{figure}

The effect of the phase  on the estimation degree of the weight parameter angle $\theta_a$ is shown in Fig.(7), where  two cases are considered of the phase, $\phi=0, \frac{\pi}{2}$ in Figs.(7a), (7b), respectively, and it is assumed that the triggers are prepared be setting $\tilde\theta_a=\tilde\theta_b=\frac{\pi}{4}$. In general, the behavior of $\mathcal{F}_{\theta_a}$ is similar to that displayed in Fig.(6a). However, the maximum values that are predicted in Fig.(7) are much smaller than those displayed in Fig.(6a).  Also, from this figure one may explore that when the initial states of the  triggers and the qubits are different, the possibility of estimating the weight parameter decreases. On the other hand, as, $\theta_b$ increases, $\mathcal{F}_{\theta}$ increases, but the increasing rate is larger than that displayed in Fig.(6a), where $\mathcal{F}_{\theta}$ increases gradually.  Fig.(7a), displays the effect of larger values of the phase, where we set $\phi=\frac{\pi}{2}$.  The behavior of the quantum Fisher information is similar to that displayed in Fig.(7a), but the upper bounds are smaller. Moreover, the increasing rate (as $\theta_b$ increases) and the decreasing rate as ($\theta_a$ increases) are smoothly compared with those shown in Fig.(7a).\\

From Fig.(6) and (7), one  may conclude that the possibility of maximizing the estimation degree, depends on the initial state of the triggers. The larger bounds are predicted when the qubits and their triggers are polarized in the same direction. The phase parameter plays the control parameter on the whole process, where the maximization of the quantum Fisher information is predicted at $\phi=0,\pi$ and $2\pi$. Otherwise it is minimized. The minimum values are affected by the initial state preparation of Alice and Bob.

\begin{figure}[!htb]
\begin{center}
\includegraphics[scale=.60]{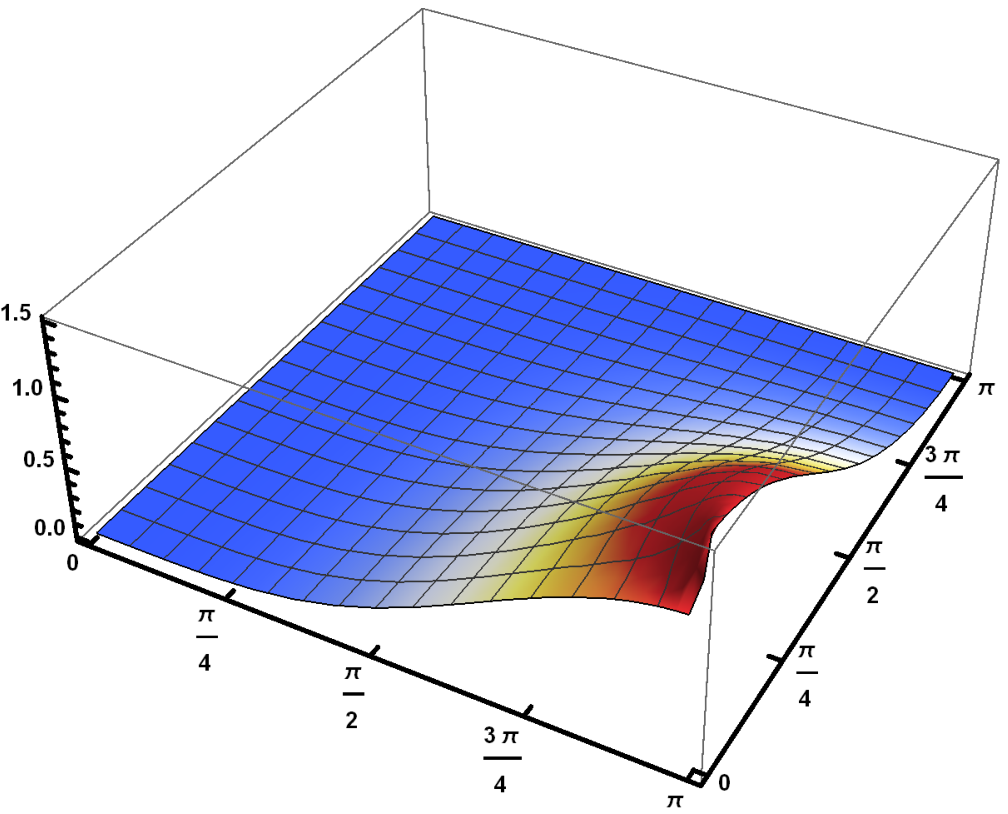}
\put(-120,150){$(a)$}
\put(-190,70){$\mathcal{F_{\theta}}_b$}
                      \put(-130,20){$\theta_{a}$}
                      \put(-20,30){$\theta_{b}$}~\quad\quad\quad
\includegraphics[scale=.60]{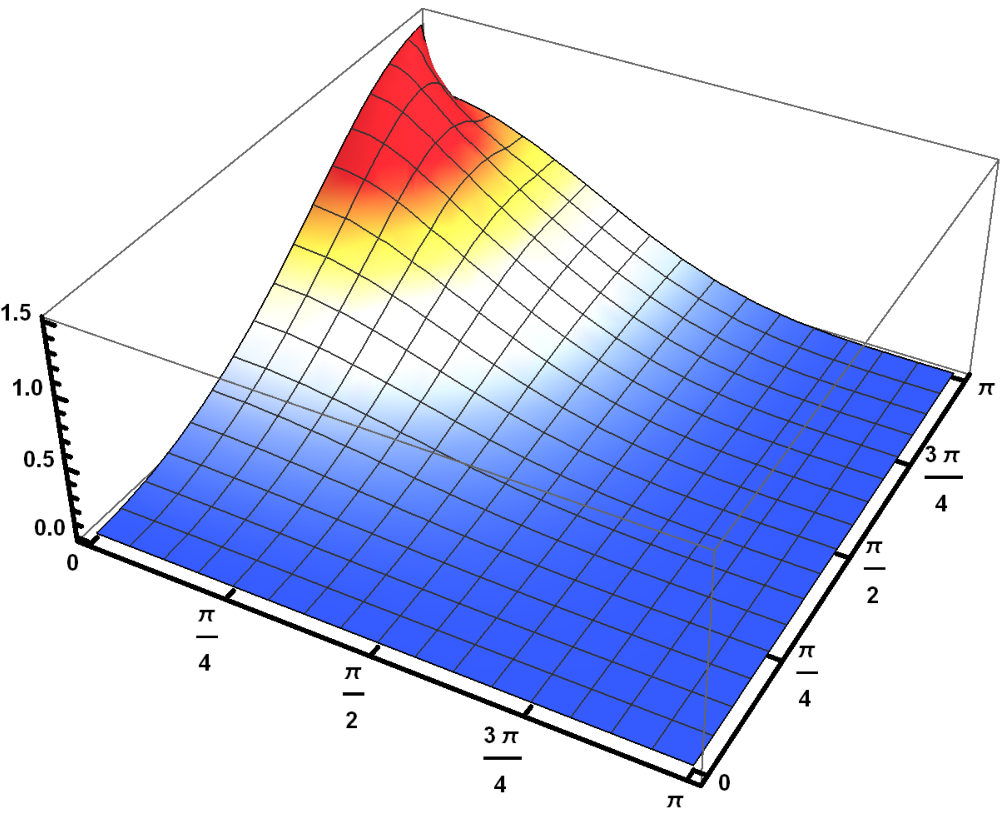}\put(-120,150){$(b)$}
\put(-190,70){$\mathcal{F_{\theta}}_b$}
                      \put(-130,20){$\theta_{a}$}
                      \put(-20,30){$\theta_{b}$}~\quad\quad\quad
\includegraphics[scale=.40]{legend.eps}\\

\includegraphics[scale=.60]{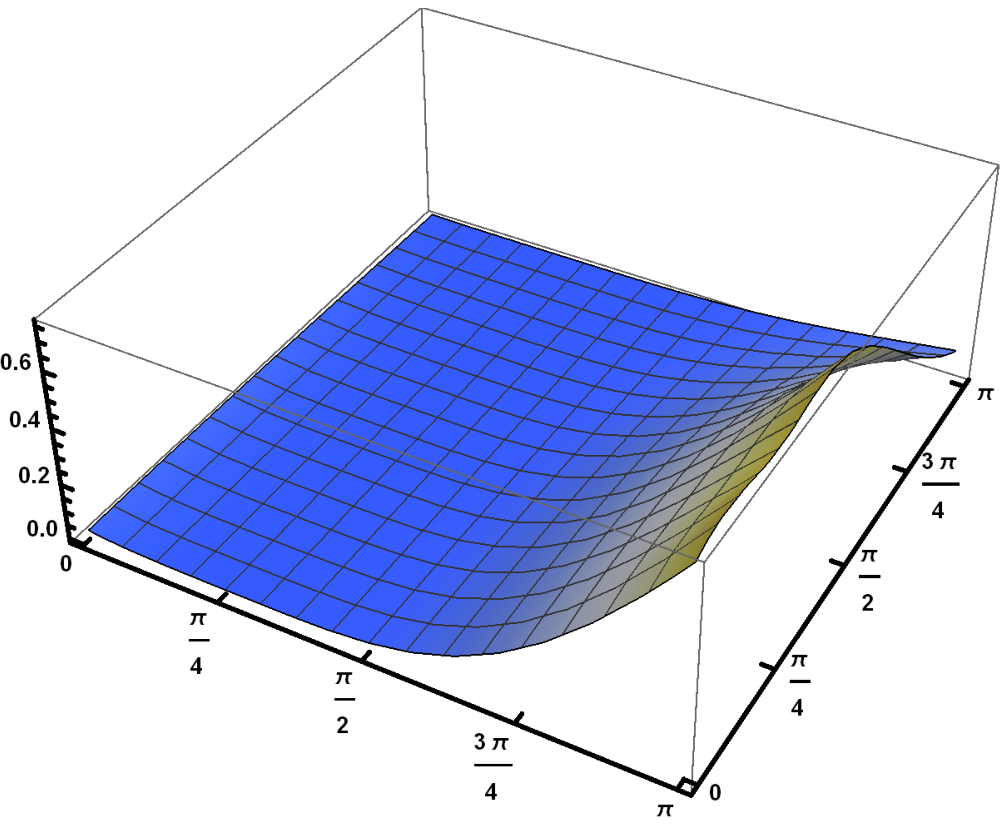}
\put(-120,150){$(c)$}
\put(-190,70){$\mathcal{F_{\theta}}_b$}
                      \put(-130,20){$\theta_{a}$}
                      \put(-20,30){$\theta_{b}$}~\quad\quad\quad
\includegraphics[scale=.60]{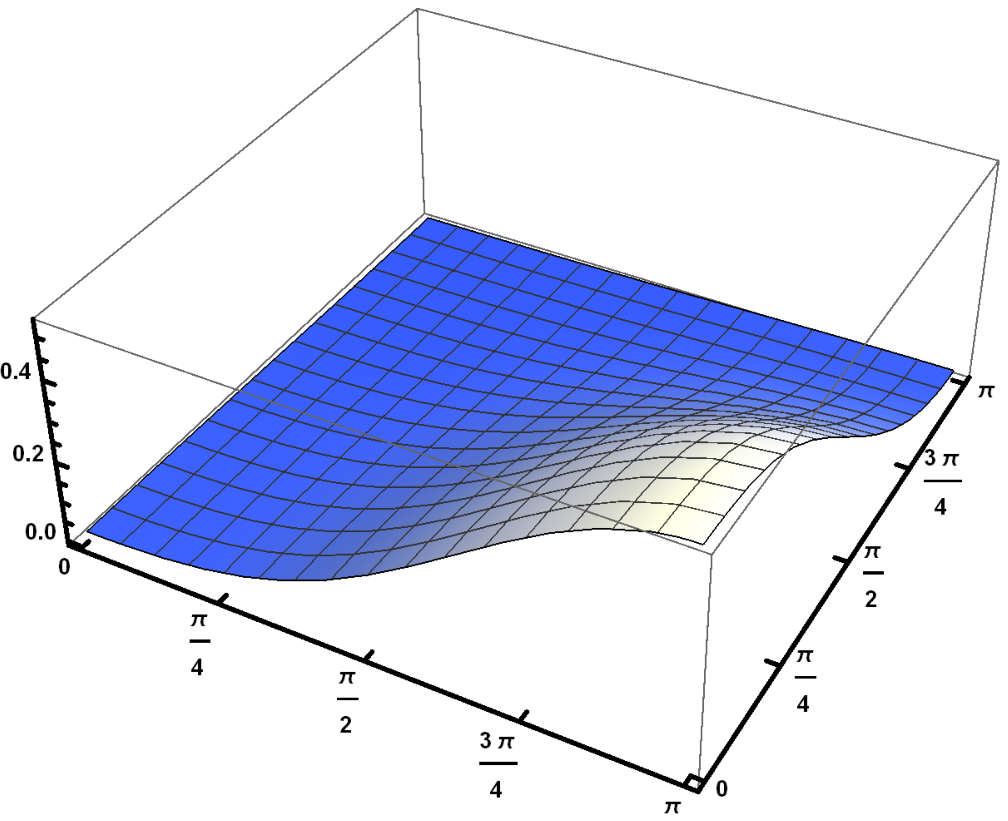}\put(-120,150){$(d)$}
\put(-190,70){$\mathcal{F_{\theta}}_b$}
                      \put(-130,20){$\theta_{a}$}
                      \put(-20,30){$\theta_{b}$}~\quad\quad\quad
\includegraphics[scale=.40]{legend.eps}
\caption{Fisher information $\mathcal{F}_{\theta_b}$ , where (a) and (b) are the same as Fig. (6), while (c) and (d) are the same as Fig.(7). }
\label{Figure1}
              \end{center}
\end{figure}

In Fig.(8), we summarize the effect of the initial state settings of Alice and Bob's qubit. As well as, the state of their triggers on the behavior of $\mathcal{F}_b$.  In Fig.(8a), we used to assume that the initial states of the trigger are prepared in the states $\frac{1}{2}(\hat{I}+\hat\sigma^{(a)}_z)$ for Alice's trigger, while Bob's trigger is prepared in the state $\frac{1}{2}(\hat{I}+\hat\tau^{(b)}_z)$, where the phase of all the qubits is arbitrary. The behavior is similar to that displayed in Fig.(6), but as the $\mathcal{F}_{\theta_a}$ increases, $\mathcal{F}_{\theta_b}$ decreases.  This means that, as the information at Alice's hand starts to increases, the information at Bob's hand decreases. Therefore the maximum bounds are predicted at different initial state settings. However, the states of  the triggers paly the same role as it is played in Fig.(6). As it is shown in Fig.(8b), the maximum bound of $\mathcal{F}_{\theta_b}$ is shown at $\theta_b=\pi$, namely $\rho_q^{(b)}=\frac{1}{2}(\hat{I}-\hat\tau^{(b)}_z)$. This means that the Bob state is prepared on its trigger's state, where $\rho_T^{(b)}=\frac{1}{2}(\hat{I}-\hat\tau^{(b)}_z)$.  On the other hand, the minimum values of the quantum Fisher information $\mathcal{F}_{\theta_b}$ are predicted at $\theta_b=0$, namely  $\rho_q^{(b)}=\frac{1}{2}(\hat{I}+\tau^{(b)}_z)$ and $\rho_q^{(a)}=\frac{1}{2}(\hat{I}-\hat\sigma^{(b)}_z)$.\\

Figs.(8c) and (8d) describe the effect of the phase on the estimation degree of $\mathcal{F}_{\theta_b}$, where we assume that the trigger states are prepared with a weight angle $\tilde\theta_a=\tilde\theta_b=\frac{\pi}{4}$.   It is clear that the behavior of the quantum Fisher information is similar to that shown of $\mathcal{F}_{\theta_a}$ as displayed in Fig.(7). However, the maximum bound of $\mathcal{F}_{\theta_b}$ is displayed at $\theta_b=\frac{\pi}{4}$, namely when the weight angle $\theta_b=\tilde\theta_b=\frac{\pi}{4}$.  Moreover at $\phi_a=\phi_b=\phi=0$, the losses on Fisher information due to the difference between the initial state preparation of both qubits.  In fact, an additional loss of the quantum Fisher information due to the existence of the phase, where at $\phi=\frac{\pi}{2}$, the maximum bounds of $\mathcal{F}_{\theta_b}$  are smaller than those shown at $\phi=0$.

\subsection{Multiparameter estimation }

The quantum Fisher information matrix (QFIM) provides a tool to determine the precision in the multiparameter estimation. Where the lower bound of the estimation is given by the Cramer Rao inequality \cite{paris} as:
\begin{equation}\label{5}
Cov(\hat\theta)  \geq  \mathcal{F}M^{-1},
\end{equation}
with $\hat\theta =\theta_1 , \theta_2 ...$ is a set of parameters can be encoded in the density matrix $\rho = \rho(\hat\theta)$  and $Cov(\hat\theta)$ is the covariance of the parameters $\hat\theta$, $\mathcal{F}M^{-1}$ is the quantum Fisher information matrix.\\
 The inverse of the QFIM gives the lower error limit of the estimation. The expression of the QFIM \cite{paris} is defined as:
\begin{equation}\label{6}
\mathcal{F}M_{ij}=Tr[\rho \frac{L_i L_j+ L_j L_i}{2}]= Tr[\partial_{i}\rho L_i]= Tr[\partial_{j}\rho L_j],
\end{equation}
with $L_i$ is the symmetric logarithmic derivative corresponds to the parameter $\theta_i$  which is determined by the equation:
\begin{equation}\label{7}
\partial_{i}\rho= \frac{1}{2}(L_i \rho + \rho L_i),
\end{equation}
using the Eqs. (\ref{6}) and (\ref{7}), two expressions of the QFIM can be derived. The first is based on the calculation of the exponential of $\rho$ and an integral as:
\begin{equation}
\mathcal{F}M_{ij}=Tr[\partial_{i}\rho L_i]= 2\int_{0}^{\infty} Tr[dt e^{-\rho t} \partial_{i}\rho e^{-\rho t} \partial_{i}\rho].
\end{equation}
The second expression to calculate the QFIM is based on the decomposition of $\rho$ into eigenvalues $ p_n$ and eigenvectors $\ket{\psi_n}$ as:
\begin{equation}
\rho= \sum_{n}  p_n \ket{\psi_n} \bra{\psi_n}
\end{equation}
\begin{equation}
\mathcal{F}M_{ij}=2\sum_{p_n + p_m > 0} { \bra{\psi_m} \partial_i\rho \ket{\psi_n} }   \bra{\psi_n} \partial_j\rho \ket{\psi_m}/ {p_n + p_m}
\end{equation}
A simple expression to calculate the QFIM is given in \cite{20}. For an invertible density matrix $\hat\rho$, the QFIM can be computed analytically as :
\begin{equation}
\mathcal{F}M_{ij}=2 Vec^{\dagger} {[\partial_i \hat \rho]} (\hat \rho^T \otimes \hat I + \hat I \otimes \hat \rho)^{-1} Vec[\partial_j \hat \rho],
\end{equation}
where $\partial_{i}=\partial/\partial_{i}$, $\hat{I}$ is the identity matrix and $\rho^{T}$ is the transpose of $\rho$. $Vec[\partial_i \hat \rho]$  is the Vector column of the matrix $\partial_i \hat \rho$.
In an explicit form, one can write the matrix representation of the quantum Fisher information matrix with respect to the two parameters $\theta_a$ and $\theta_b$ as,
\begin{equation}
\mathcal{F}_{\theta_a ,\theta_b}=
\begin{pmatrix}
\mathcal{F}_{\theta_a ,\theta_a} & \mathcal{F}_{\theta_b ,\theta_a}\\
\mathcal{F}_{\theta_a ,\theta_b} & \mathcal{F}_{\theta_b ,\theta_b}
\end{pmatrix},
\end{equation}
where,
\begin{equation}\label{BlochQFI}
\mathcal{F}_{\theta_i,\theta_j} = \left\{
\begin{array}{ll}
(\partial_{\theta_i} \overrightarrow{a})(\partial_{\theta_j} \overrightarrow{a})+\frac{(\overrightarrow{a}.\partial_{\theta_i}\overrightarrow{a})(\overrightarrow{a}.\partial_{\theta_j}\overrightarrow{a})}{1-|\overrightarrow{a}|^{2}},  & \mbox{if } |\overrightarrow{a}|<1 \\
(\partial_{\theta_i} \overrightarrow{a})(\partial_{\theta_j} \overrightarrow{a}), & \mbox{if }
|\overrightarrow{a}|=1.
\end{array}
\right.
\end{equation}
Explicitly on Alice side the QFIM is given by,
\begin{equation}\label{Alice}
\mathcal{F}_{\theta_a ,\theta_b}^{Alice}=\frac{1}{1-(p_b \bar p_a)^2}
\begin{pmatrix}
(p_b \partial_{\theta_a} \bar p_a)^2  & p_b \bar p_a \partial_{\theta_b}p_b \partial_{\theta_a}\bar p_a \\
\\
 p_b \bar p_a \partial_{\theta_b}p_b \partial_{\theta_a}\bar p_a & ({p_b\bar p_a})^2(1-(p_b \bar p_a)^2) +(\bar p_a \partial_{\theta_b}p_b)^2
\end{pmatrix}.
\end{equation}
A similar expression is obtained for QFIM at Bob's side, just we switched between the suffix $a$ and $b$. However, the amount of Fisher  information  at  the users Alice (Bob)  that injected on the parameters $\theta_b(\theta_a)$  respectively is   given by,
\begin{equation}
\mathcal{F}_{\theta_b}=\frac{ ({p_b\bar p_a})^2(1-(p_b \bar p_a)^2) +(\bar p_a \partial_{\theta_b}p_b)^2}{1-(p_b \bar p_a)^2}
,\quad\quad
\mathcal{F}_{\theta_a}\frac{ ({p_a\bar p_b})^2(1-(p_a \bar p_b)^2) +(\bar p_b \partial_{\theta_a}p_a)^2}{1-(p_a \bar p_b)^2}.
\end{equation}
By using the Cramer-Rao bound for one parameter estimation, one can estimate the single parameter by means of the variance as,
\begin{equation}
Var(\theta_a)^{Ind} \geq \frac{1-(p_a \bar p_b)^2}{({p_a\bar p_b})^2(1-(p_a \bar p_b)^2) +(\bar p_a \partial_{\theta_a}p_a)^2}\quad\quad;Var(\theta_b)^{Ind} \geq \frac{1-(p_b \bar p_a)^2}{({p_b\bar p_a})^2(1-(p_b \bar p_a)^2) +(\bar p_a \partial_{\theta_b}p_b)^2},
\end{equation}
where it is assumed that the estimation process of each parameter is  independent. However, the variance estimation of the two parameters $\theta_a$ and $\theta_b$  is given  simultaneously as \cite{23},
\begin{equation}
Var(\theta_b)^{Sim} \geq \frac{(p_b \partial_{\theta_b} \bar p_a)^2}{\bar p_a ^2 p_b^4 (\partial_{\theta_a} \bar p_a   )^2}\quad\quad;Var(\theta_a)^{Sim} \geq \frac{(p_a \partial_{\theta_b} \bar p_b)^2}{\bar p_b ^2 p_a^4 (\partial_{\theta_b} \bar p_b   )^2}.
\end{equation}
The minimum variance of the two parameters my be described by,
\begin{equation}
\Delta=\frac{V_I(\theta_a)}{V_S(\theta_a)},~ \quad
\delta=\frac{V_I(\theta_b)}{V_S(\theta_b)}.
\end{equation}

\begin{figure}[!htb]
\begin{center}
\includegraphics[scale=.70]{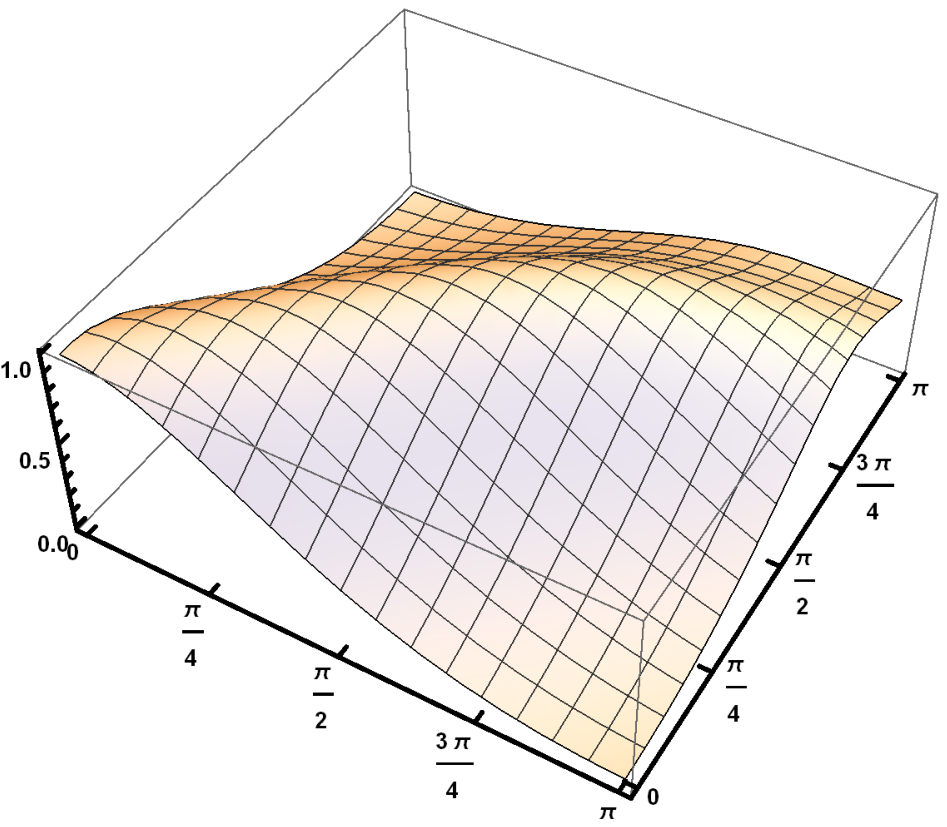}
\put(-200,85){$\Delta$}
                   \put(-120,170){$(a)$}   \put(-140,20){$\theta_{a}$}
                      \put(-20,40){$\tilde\theta_{t}$}~~~\quad\quad\quad
                      \includegraphics[scale=.50]{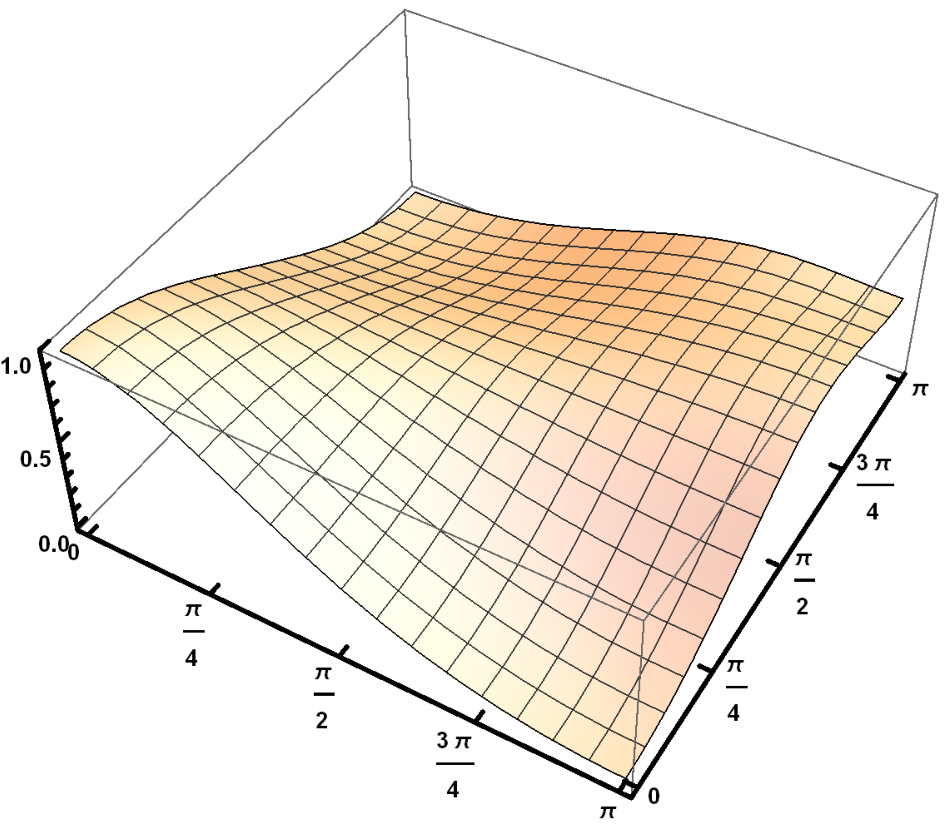}
\put(-195,85){$\delta$}
                   \put(-120,170){$(b)$}   \put(-140,20){$\theta_{b}$}
                      \put(-20,40){$\tilde\theta_{t}$}~~~\quad\quad\quad
\caption{Ratios  of estimations $\Delta$ (for Alice) with  $\theta_b=\frac{\pi}{4}$ and $\delta$ (for Bob), with  $\theta_a=\frac{\pi}{4}$, where we set  $\tilde\theta_a=\tilde\theta_b =\tilde\theta_t$ , where we set $\phi = 0$ for  (a,c) and $\phi$ = $\frac{\pi}{2}$ for  (b,d)}
\label{Figure2}
              \end{center}
\end{figure}

In Fig.(9)  we investigate  the behavior of the ratio variances  $\Delta$ and $\delta$ for different initial state settings.  It is clear that the ratios $\Delta<1$, and  $\delta<1$ namely   $V_{S}{(\theta_i)}>V_{I}{(\theta_i)}, ~i=a,b$ and consequently those ratios satisfy, the maximum/minimum probability  of quantum mechanics, where  $\Delta$ and $\delta\in[0,~1]$. So, the ratios demonstrate the full classical/quantum systems at $0$ and $1$, respectively.
Moreover, these ratios obey the uncertainty principle for $0\leq\Delta(\delta)\leq 1$. On the  other hand, for individual estimation, we can see that the quantum Fisher information  ($\mathcal{F}_{\theta_a}(  \mathcal{F}_{\theta_b)})>1$, which outside the range of entangled systems. Therefore, using the multi-parameter form is much better than the  single parameter form.
Fig.(9a), describes the behavior of $\Delta(\theta_a,\tilde\theta_t)$ at a particular initial state of Bob's qubit, where we set $\theta_b=\frac{\pi}{4}$,with  zero-phase for all the qubits. Also, the maximum/minimum bounds are depicted when the qubits and its trigger have the same polarization. As an example, the maximization is predicted at $\theta_a=\tilde\theta_t=0$ or $\pi$, while the minimum values are predicted when $\theta_a$ and $\tilde\theta_t$ are different.  The same behavior is predicted for the ratio $\delta$, but the maximum and minimum are displayed at different angles.

\section{Conclusion}\label{section4}

In this contribution, we are interested in bidirectional quantum Fisher information between  two users, Alice and Bob. Since the form of the Fisher information is based on Bloch vectors, therefore we reformulated the bidirectional protocol of Kiktenko in terms of Bloch vectors. It is assumed that, the users' qubits and their trigger's qubits are described by using the Bloch vectors. Moreover, the local operations of CNOT and CCNOT gates are represented by Pauli-operators and the circuit which describes this protocol is introduced clearly.\\

We have discussed the fidelity of the bidirectional teleported states between the two users, where analytical forms of the fidelity are obtained. These fidelities depend on the initial states of the teleported qubit as well as on the states of the triggers. It is shown that, the fidelity of the teleported state is maximized  when the teleported qubit and its trigger are polarized  in the same direction.  However, the initial state settings of the qubits play an important role in maximizing/minimizing the fidelity of the teleported state, where the maximization is achieved if both qubits are initially prepared in the same direction. The minimization takes place at non-zero phase  angle or different initial state settings. The fidelity of the teleported  classical information  bidirectionally over the quantum channel is much better than teleporting quantum information.\\

The possibility of estimating the weight parameters of the teleported states bidirectionally is quantified by using quantum Fisher information. An analytical form based on the Bloch vectors is introduced.  Similarly,  the fidelity depends on the initial states of all the qubits.
It is shown  that the possibility of  maximizing the estimation degree depends on the initial state of the triggers. The larger bounds are predicted when the qubits and their triggers are polarized in the same direction. The phase parameter plays the control parameter on the whole process, where the maximization of the quantum Fisher information is predicted  at $\phi=0, \pi,2\pi$,  otherwise it is minimum. The minimum values are affected by the initial state settings  of Alice and Bob.\\

In addition to estimating  a single parameter, we quantify the weight parameters of the bidirectional states by means of the variances' ratios.
It is shown that the multi-parameter estimation technique is much better than the single parameter estimation, where the ratios running between "0" for fully classical systems  and "1" for completely entangled system as well  as the uncertainty principle is obeyed.
It is shown that, the maxim/minimum estimation of these parameters depends on the initial states of the triggers.  Also, as the maximum estimation is predicted at Alice 's side, the minimum estimation is displayed at Bob's side.\\

Since, the initial states of Alice and Bob, as well as the states of their trigger are similar, therefore the maximum/minimum values of the fidelity of the teleported states  and  the  quantum Fisher information are the same  for both directions. However, as soon as Alice got the maximum information, Bob lost all the information and vice versa. Therefore, the maximum bounds of the fidelity and the quantum Fisher information are obtained at different weight angles.

\end{document}